\documentclass[twocolumn, superscriptaddress, reprint, amsmath,amssymb, floatfix, aps, prd]{revtex4-2}

\usepackage{amsmath,amssymb}
\usepackage[english]{babel}
\usepackage{graphicx}
\usepackage{subcaption}
\usepackage{xcolor}
\usepackage{float}
\usepackage{braket}
\usepackage{bm}
\usepackage{array}

\begin{document}

\title{Accurate Unsupervised Photon Counting from Transition-Edge Sensor Signals}
\author{Nicolas Dalbec-Constant}
\email{nicolas.dalbec-constant@polymtl.ca}
\affiliation{D\'epartement de g\'enie physique, \'Ecole Polytechnique de Montr\'eal, Montr\'eal, QC, H3T 1J4, Canada}

\author{Guillaume Thekkadath}
\affiliation{National Research Council Canada, 100 Sussex Drive, Ottawa, Ontario K1N 5A2, Canada}
\author{Duncan England}
\affiliation{National Research Council Canada, 100 Sussex Drive, Ottawa, Ontario K1N 5A2, Canada}
\author{Benjamin Sussman}
\affiliation{National Research Council Canada, 100 Sussex Drive, Ottawa, Ontario K1N 5A2, Canada}
\author{Thomas Gerrits}
\affiliation{National Institute of Standards and~Technology, 100 Bureau Drive, Gaithersburg, MD 20899, USA}
\author{Nicolás Quesada}
\email{nicolas.quesada@polymtl.ca}
\affiliation{D\'epartement de g\'enie physique, \'Ecole Polytechnique de Montr\'eal, Montr\'eal, QC, H3T 1J4, Canada}

\begin{abstract}
	We compare methods for signal classification applied to voltage traces from transition-edge sensors (TES) which are photon-number resolving detectors fundamental for accessing quantum advantages in information processing, communication and metrology. We quantify the impact of numerical analysis on the distinction of such signals. Furthermore, we explore dimensionality reduction techniques to create interpretable and precise photon-number embeddings. We demonstrate that the preservation of local data structures of some nonlinear methods is an accurate way to achieve unsupervised classification of TES traces. We do so by considering a confidence metric that quantifies the overlap of the photon-number clusters inside a latent space. Furthermore, we demonstrate that for our dataset previous methods such as the signal's area and principal component analysis can resolve up to 16 photons with confidence above $90\%$ while nonlinear techniques can resolve up to 21 with the same confidence threshold. Also, we showcase implementations of neural networks to leverage information within local structures, aiming to increase confidence in assigning photon numbers. Finally, we demonstrate the advantage of some nonlinear methods to detect and remove outlier signals.
\end{abstract}

\maketitle

\section{Introduction}
Photonics is a strong contender for building large-scale quantum information processing systems~\cite{arrazola_quantum_2021,slussarenko_photonic_2019,rudolph2017optimistic,bourassa2021blueprint,maring2024versatile}; in many of these systems, photon-number detection plays an essential role, serving as a resource for quantum advantage. Photonic architectures often encode information in continuous variables or in multi-photon states, where precise knowledge of photon number is critical for state preparation, measurement, and error correction. For example, photon-number-resolving detectors have been used for the generation of non-Gaussian states~\cite{takase2024generation,alexander2024manufacturable,yao2022design,chen2024generation,melalkia2023multiplexed,tiedau2019scalability,sonoyama2024generation,endo2024optically,gerritsGenerationOpticalCoherentstate2010,thekkadathEngineeringSchrodingerCat2020, larsenIntegratedPhotonicSource2025}, for the sampling of classically-intractable probability distributions~\cite{aaronson2011computational,hamilton2017gaussian,kruse2017limits,deshpande2022quantum,grier2022complexity,madsen2022quantum}, or for directly resolving multiple quanta, thereby improving the Fisher information in interferometric protocols~\cite{thekkadath2020quantum, Wildfeuer:09,youScalableMultiphotonQuantum2021}. In such protocols, quantum states of light such as N00N or Holland-Burnett states can, in principle, achieve a precision scaling of $\sqrt{N}$ (for $N$ detected photons), surpassing classical limits. However, this quantum advantage is highly sensitive to loss, making efficient photon-number resolution particularly valuable. The use of photon-number resolving detectors offers a significant advantage, as a single device can accurately determine the number of photons in a quantum state~\cite{divochiy_superconducting_2008,moraisPreciselyDeterminingPhotonnumber2022a}, thereby eliminating the need for a demultiplexed network of threshold detectors, which adds complexity and can reduce overall efficiency~\cite{kruse2017limits,jonsson2019evaluating,jonsson2020temporal}. Transition-edge sensors (TES) have been used for this task, offering resolution over a wide energy range. Resolutions up to 30 photons have been demonstrated~\cite{eaton2023}, although this quantity is typically lower, on the order of 17~\cite{moraisPreciselyDeterminingPhotonnumber2022a}.

TESs exploit the superconducting phase transition of photosensitive materials (illustrated in Fig.~\ref{fig:circuit}) to achieve an extremely sensitive calorimeter~\cite{irwin_transition-edge_2005}. During operation, the material is cooled below its critical temperature and then current-biased to the transition region between its superconducting and normal state. In this region, the temperature increase following the absorption of a single photon leads to a measurable change in the material's resistance~\cite{phillips2020advanced,hadfield2009single}. The resistance change is read-out using a low noise amplifier such as superconducting quantum interference devices (SQUIDs), which also enable the creation of large arrays of TES detectors via read-out multiplexing~\cite{irwin_transition-edge_2005}. Optimized materials and coupling techniques have demonstrated system efficiencies of up to 98\%~\cite{fukuda_titanium-based_2011}.

The readout of these devices is non-trivial as the quantity one wants to determine, the energy (or the photon number for monochromatic light), is reflected in a nonlinear fashion in the voltage signal produced by the detectors' electronics~\cite{gerrits_extending_2012}. Historically, the integral (area) of the signals has been used to assign photon numbers \cite{moraisPreciselyDeterminingPhotonnumber2022a,Schmidt_Bimodal_2021}. However, distinguishing large photon numbers becomes challenging with this technique. To address this issue, linear techniques such as Principal Component Analysis (PCA) have been used~\cite{humphreys_tomography_2015}. A machine learning method, adapted from the K-means algorithm to account for the Poissonian statistics of laser sources, has also been developed~\cite{levine_algorithm_2012}. However, these methods' simplicity or assumptions can limit their performance or usability for model-free photon-number detection and when measuring non-classical sources, which typically do not have Poisson photon-number statistics.

With the increased popularity of machine learning in the field of signal processing~\cite{rajendran_deep_2018} and quantum systems~\cite{nautrup_optimizing_2019}, one might naturally ask whether employing more sophisticated methods could lead to enhanced resolution of photon numbers. In this work, we answer this question by assessing the performance of multiple techniques for photon number classification using TES signals. We do so by considering a confidence metric that quantifies the overlap of the photon-number clusters inside a latent space. We demonstrate that for our dataset previous methods such as the signal's area and PCA can resolve up to 16 photons with confidence above $90\%$ while nonlinear techniques can resolve up to 21 with the same confidence threshold. Furthermore, we also showcase implementations of neural networks to leverage information within local structures, aiming to increase confidence in assigning photon numbers. Finally, we demonstrate the advantage of some nonlinear methods to detect and remove outlier signals.

Our manuscript is structured as follows: in the next section, Sec.~\ref{sec:methodology}, we formulate the problem of photon-number discrimination in the general setting of unsupervised classification and dimensionality reduction. Next, in Sec.~\ref{sec:methods}, we offer a brief overview of the methods used to compute similarities between signals and how we distinguish signals that belong to the different photon-number classes. We present our results in Sec.~\ref{sec:results} using experimental data, followed by a discussion of the use cases of the described methods in Sec.~\ref{sec:discussion}

\section{Methodology}\label{sec:methodology}
\subsection{Problem Formulation}
Consider a data matrix $\bm{X}\in \mathbb{R}^{u \times t}$ that stores $u$ signals $x_i$ of size $t$. We assume there exists an operation $f(\bm{X})$ that can transform $\bm{X}$ into a vector $\bm{n}\in \mathbb{R}^{u\times 1}$ that contains the photon number associated with every signal. The goal of the classification becomes finding a parametric transformation $F(\theta', \bm{X})$ with user-defined parameters $\theta'$ that approximates as closely as possible the true transformation $f(\bm{X})$.

The problem is defined as an unsupervised classification, meaning the true elements of $\bm{n}$ are unknown. Additionally, given an experiment, the method needs to accept arbitrarily high photon numbers within the visibility limit of the detector.

\subsection{Dimensionality Reduction}
To solve this unsupervised classification problem, dimensionality reduction techniques are used. This process describes the transformation of $\bm{X}$ into a lower-dimensional output $\bm{Y}\in \mathbb{R}^{u\times r}$ that retains a meaningful amount of the input information. The new space of dimension $r<t$ is referred to as a latent space and is limited to one and two dimensions in this study. The proposed approach could be used for an arbitrarily large latent space, although these higher dimensional spaces are harder to interpret.

We use dimensionality reduction since it is a natural extension of previous work that uses PCA~\cite{humphreys_tomography_2015}. Moreover, this framework is used to make the current work compatible with existing tomography routines~\cite{humphreys_tomography_2015}.
It also enables the visualization and interpretation of an entire dataset, a task difficult by directly observing the TES signals. Supposing an accurate transformation exists and is faster to process than the acquisition rate of the detector, the low-dimensional representation reduces the memory requirements of experiments by acting as a compression step.
\begin{figure*}[t]
	\centering
	\begin{subfigure}[t]{0.45\textwidth}
    	\centering
    	\includegraphics[width=\textwidth]{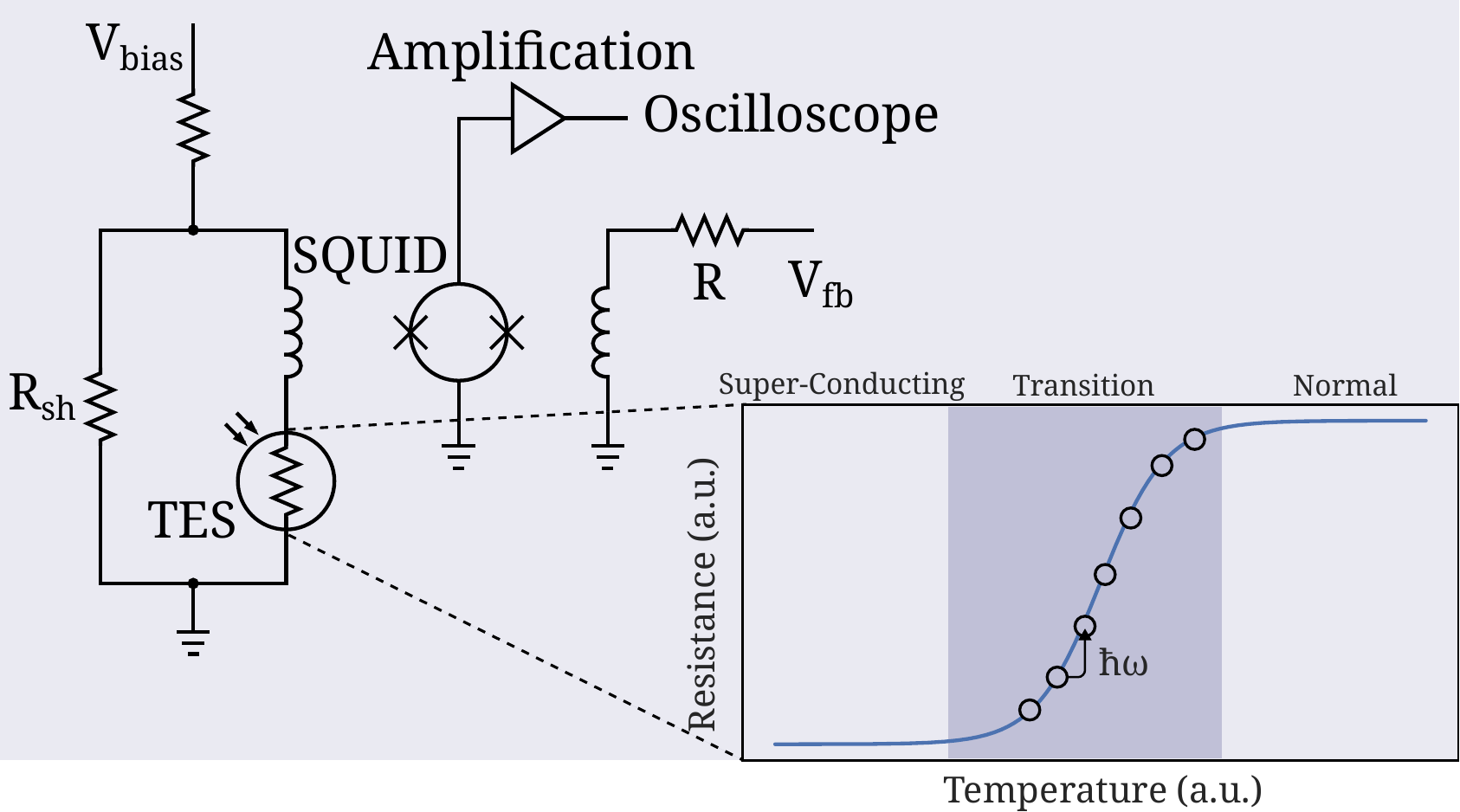}
    	\caption{Circuit diagram of a typical transition-edge sensor detection scheme. The circuit can change slightly from one implementation to the other, the illustrated circuit is based on Ref.~\cite{moraisPreciselyDeterminingPhotonnumber2022a}. The superconducting phase transition of the TES is illustrated through the sharp variation of resistance as a function of temperature, giving the TES extremly sensitive energy resolution~\cite{thekkadathPreparingCharacterizingQuantum2020b}.}
    	\label{fig:circuit}
	\end{subfigure}
	\hfill
	\begin{subfigure}[t]{0.45\textwidth}
    	\centering
    	\includegraphics[width=\textwidth]{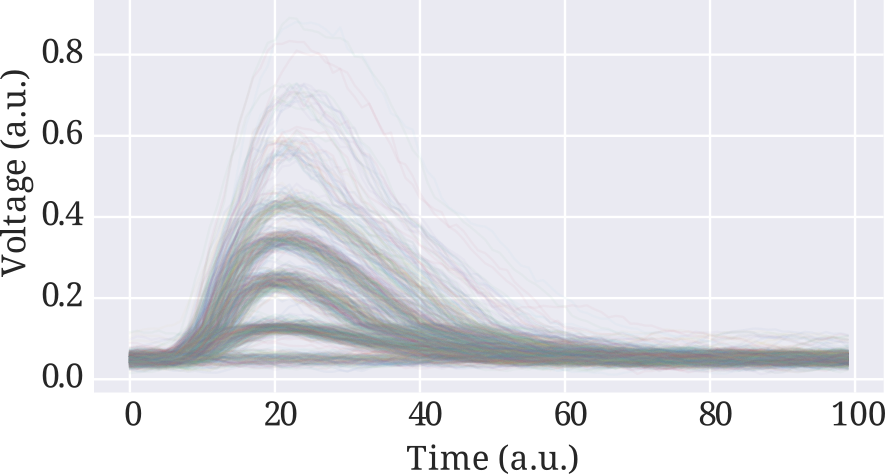}
    	\caption{Example of multiple voltage outputs measured by an oscilloscope to construct a dataset $\bm{X}$ with $u=1\,024$ raw TES traces of size $t=100$.}
    	\label{fig:EXtraces}
	\end{subfigure}
	\hfill
	\begin{subfigure}[t]{0.45\textwidth}
    	\centering
    	\includegraphics[width=\textwidth]{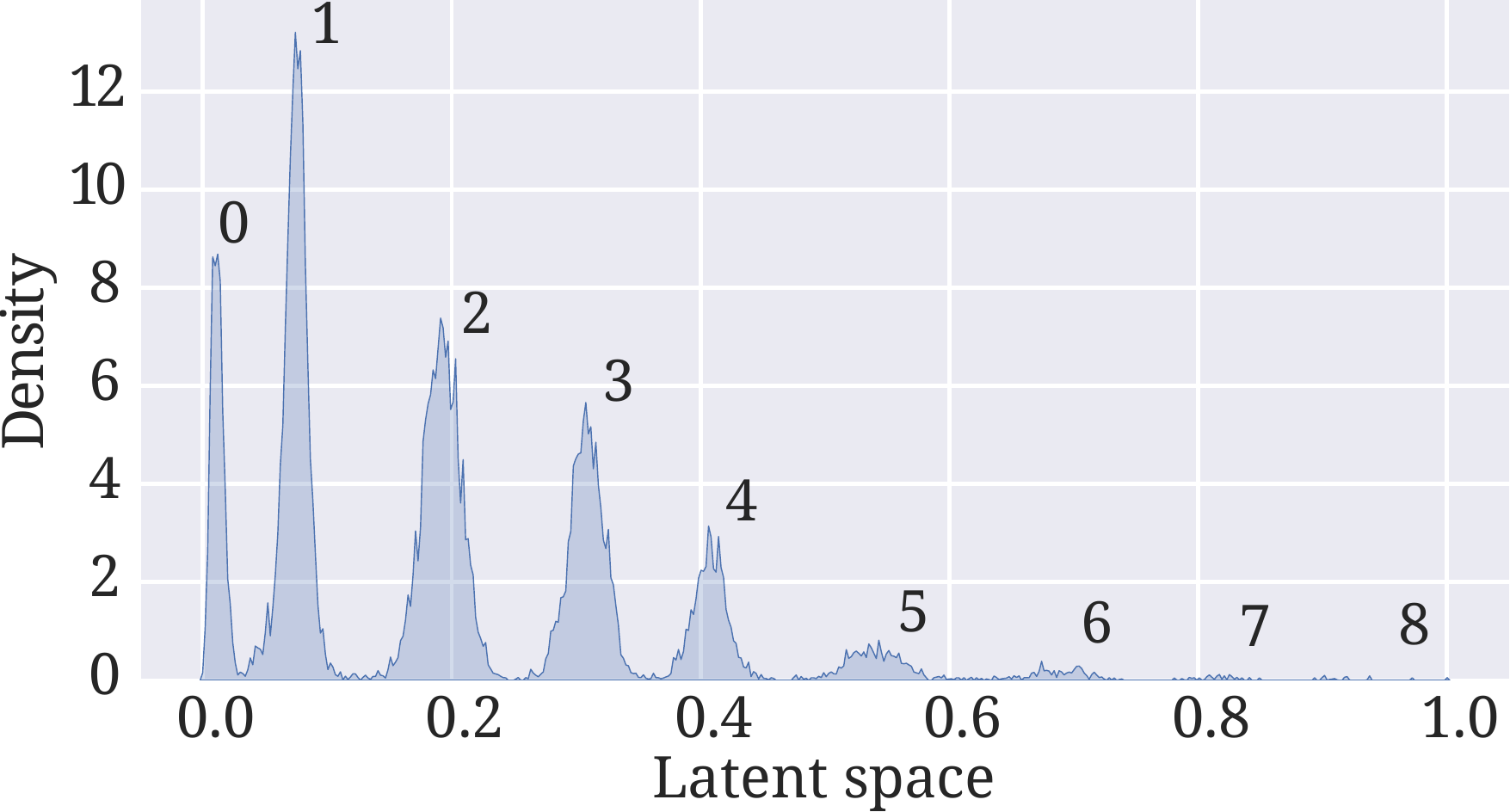}
    	\caption{The dataset $\bm{X}$ is transformed into $\bm{Y}$ which has a single dimension ($r=1$), here plotted using a kernel density estimation~\cite{SeabornkdeplotSeaborn0132}. The dimensionality reduction technique (maximum value of the signals in this case) creates a low-dimensional space where signal features become apparent. Each peak is a cluster that represents the underlying dominant feature of the signals: the photon numbers.}
    	\label{fig:EXdensity}
	\end{subfigure}
	\hfill
	\begin{subfigure}[t]{0.45\textwidth}
    	\centering
		\raisebox{1.5ex}{\includegraphics[width=\textwidth]{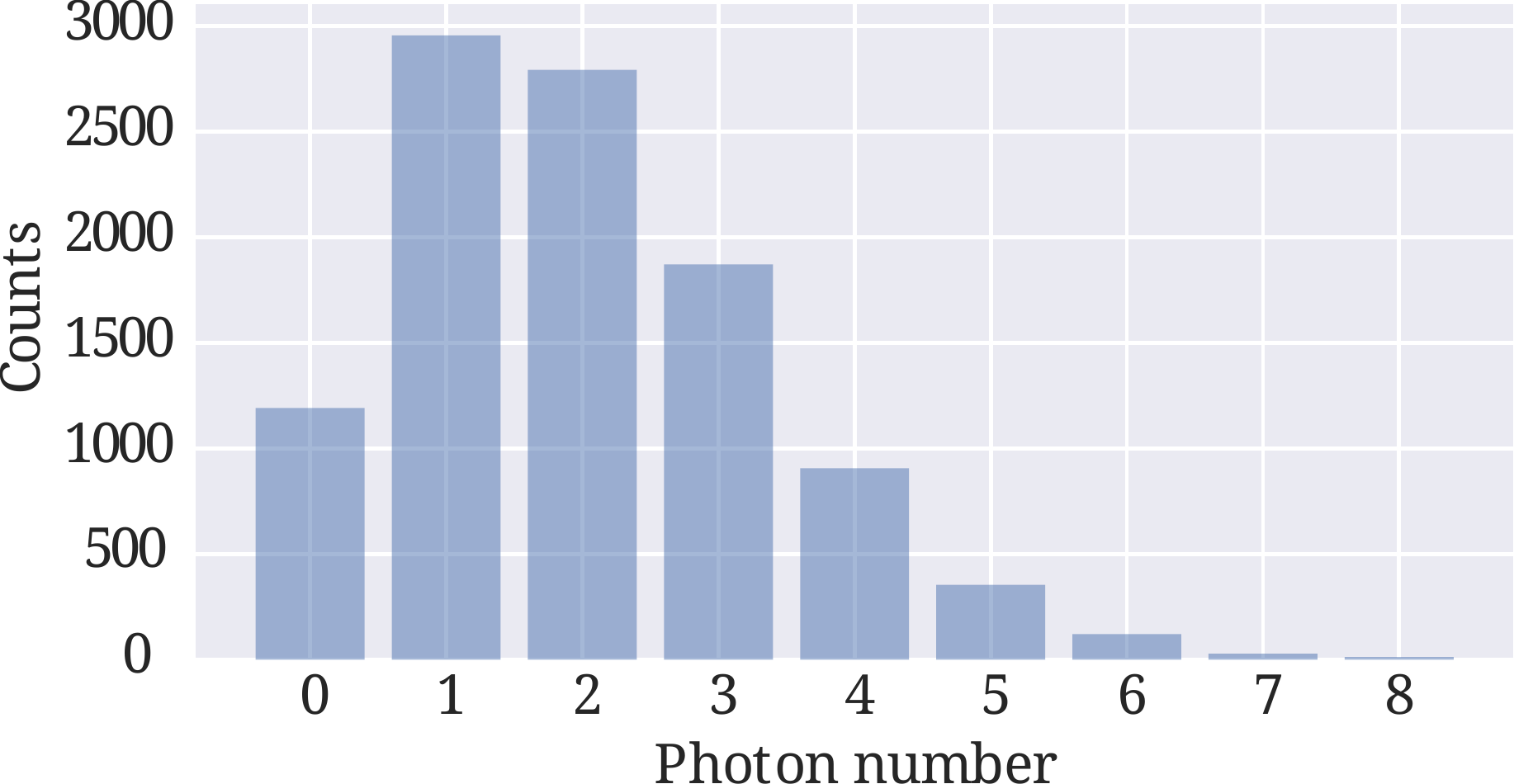}}
    	\caption{Clusters in the latent space are assigned a photon number $n\in\{0,1,...,8\}$. This is done by dividing the space in regions most likely associated to specific photon numbers (see Sec.~\ref{sec:conf}). From labelled samples, a photon-number distribution can be generated.}
    	\label{fig:EXdist}
	\end{subfigure}
	\caption{Steps associated with the photon number detection process going from the device in~\textbf{(\ref{fig:circuit})} and the expected output signals in~\textbf{(\ref{fig:EXtraces})} to the abstract space describing similarities between samples in~\textbf{(\ref{fig:EXdensity})} and the final photon number distribution associated with an experiment in~\textbf{(\ref{fig:EXdist})}.}
	\label{fig:latentSpaceEx}
\end{figure*}
Considering every signal in $\bm{X}$ can be associated with a photon number $n\in\{0,1,...,c\}$, where $c$ is the photon-number cutoff, i.e., the largest distinguishable photon number. We assume that effective dimensionality reduction organizes similar samples near each other, forming regions of high density.

We illustrate the process in Fig.~\ref{fig:latentSpaceEx} by transforming the TES signals (Fig.~\ref{fig:EXtraces}) into one-dimensional samples presented in Fig.~\ref{fig:EXdensity}. This low dimensional space is visualized using a kernel density estimation of the latent space (Gaussian kernel)~\cite{SeabornkdeplotSeaborn0132}. From the position of the samples in the latent space (never considering the density estimation in the computation) it is possible to find regions most likely to describe a photon number $n\in\{0,1,...,8\}$. We discuss this step in Sec.~\ref{sec:clustering}. Finally, from this interpretation of the low-dimensional space, a photon number can be assigned to every sample (Fig.~\ref{fig:EXdist}). The regions of high density in Fig.~\ref{fig:EXdensity} are called clusters and are associated with photon numbers. We note that clusters can be defined using other heuristics like neighbour distances.

An additional justification for the use of dimensionality reduction in combination with clustering instead of directly clustering over high dimensional data is that existing work has empirically demonstrated that creating a low dimensionality embedding increases the clustering capabilities in unsupervised settings~\cite{allaoui_considerably_2020}.

\section{Methods}\label{sec:methods}
We test a wide range of methods to showcase different approaches to the dimensionality reduction task. Due to the range of published solutions to the dimensionality reduction task, we limit our tests to the methods described in this section.

With experimental motivations, we consider the properties and use cases of dimensionality reduction techniques. To do so, the methods are divided into three categories based on their characteristics: basic features, non-predictive, and predictive.

\subsection{Basic Features}
The methods in this category rely on some feature with physical  significance, and their latent space represents the value of this feature. These methods are fast to compute due to their simplicity and can be combined with noise filtering to increase resolution~\cite{moraisPreciselyDeterminingPhotonnumber2022a}. To offer a fair representation of these techniques, we apply a Butterworth low-pass filter to the signals. We do not explore alternative representation of the traces by strongly reducing the frequency content of the signals. The filtering is only used in this case to remove high frequency electrical noise and preserve the original shape of the signals.  

\subsubsection{Maximum Value}
The maximum value of the signals has been used in some cases for photon-number resolution~\cite{moraisPreciselyDeterminingPhotonnumber2022a}. For experiments that only require the measurement of low photon numbers, sufficient information is found in the maximum value. For high enough photon numbers, the traces reach a plateau and the maximum value no longer gives information \cite{gerrits_extending_2012}.

\subsubsection{Area}
TES pulse area relates non-linearly to the energy absorbed by the sensor and therefore can be used for dimensionality reduction~\cite{moraisPreciselyDeterminingPhotonnumber2022a}. The area is sensitive to noise outside the pulse, hence filtering and background rejection are used in some cases to increase the performance of this method. In this work, we only filter the traces and set a threshold to reduce the influence of noise. Following existing work, the threshold is defined above the noise distribution in the flat region of the TES signals (see time steps 80 to 100 in Fig.~\ref{fig:EXtraces}, where only vacuum is detected)~\cite{moraisPreciselyDeterminingPhotonnumber2022a}.

\subsection{Non-Predictive Methods}
The methods in this category organize data within a latent space by considering the entire dataset. However, once computed, these methods do not provide a transformation that can be directly applied to new data. To predict the position of a new sample in the latent space, the entire dataset must be recomputed. As a result, these methods are less scalable and are better suited for post-processing data.

\subsubsection{t-Distributed Stochastic Neighbour Embedding (t-SNE)}
The method t-SNE is non-predictive and attempts to create a low-dimensional representation of the data by organizing all the samples in a smaller space. The position of the samples is assigned using a gradient descent by minimizing the Kullback-Leibler divergence ($\mathrm{KL}$)
\begin{align}\label{eq:KL}
	\mathrm{KL}(P \| Q) =\sum_{i=1}^u \sum_{\substack{j=1 \\ j \neq i}}^u p_{i j} \log \frac{p_{i j}}{q_{i j}}.
\end{align}
In the $\mathrm{KL}$ divergence, $p_{ij}$ represents the joint probabilities that describe the similarities between high-dimensional samples $x_i$ and $x_j$, and $q_{ij}$ are the joint probabilities for low-dimensional samples $y_i$ and $y_j$~\cite{maatenVisualizingDataUsing2008}. The high-dimensional joint probabilities are set to be symmetric conditional probabilities defined as
\begin{align}
	p_{ij}=\frac{p_{j|i}+p_{i|j}}{2u},
\end{align}
with conditional probabilities defined using Gaussian functions
\begin{align}
	p_{j|i} = \frac{\exp\left[ -\tfrac12 ||x_i - x_j||^2 /\sigma_i^2\right]}{\sum_{\substack{k=1\\k\neq i}}^u \exp\left[ -\tfrac12||x_i - x_k||^2 / \sigma_i^2\right]},
\end{align}
where $||x|| = (\sum_{i} x_i^2)^{1/2}$ represents the Euclidean norm. In low-dimensional space, the joint probabilities are given by the Student t-distribution
\begin{align}
	q_{ij} = \frac{(1+||y_i-y_j||^2)^{-1}}{\sum_{k=1}^u \sum_{\substack{l=1 \\ l \neq k}}^u (1+||y_k - y_l||^2)^{-1}}.
\end{align}
To offer high resolution over local structures in the data the variance $\sigma_i^2$ of each high dimensional Gaussian is tuned using an information parameter called the Perplexity. Perplexity is defined as
\begin{align}
	\text{Perp}(P_i) = 2^{H(P_i)},
\end{align}
where $H(P_i)$ is the Shannon entropy
\begin{align}
	H(P_i) = -\sum_{j=1}^u p_{j|i} \log_2 p_{j|i}.
\end{align}
Perplexity, initially introduced in the field of speech recognition, is user-defined and is often described as an effective number of neighbours~\cite{Jelinek1977PerplexityaMO}. The intuition behind this value is that the variance of each Gaussian in the high dimensional space is tuned to have a tail with a limited number of relevant neighbours. This means neighbours outside the effective range of the Gaussian will have similarity values considerably smaller.

\subsubsection{Uniform Manifold Approximation and Projection (UMAP)}
We describe UMAP by emphasizing its similarities with t-SNE. UMAP makes use of stochastic approximate nearest neighbour search and stochastic gradient descent to optimize a cross-entropy cost function~\cite{mcinnes_umap_2020} defined as
\begingroup\makeatletter\def\f@size{9}\check@mathfonts
\def\maketag@@@#1{\hbox{\m@th\large\normalfont#1}}%
\begin{align}
	C=\sum_{i=1}^u \sum_{\substack{j=1 \\ j \neq i}}^u v_{ij}\log\left(\frac{v_{ij}}{w_{ij}}\right) + (1-v_{ij}) \log\left(\frac{1-v_{ij}}{1-w_{ij}}\right),
\end{align}
\endgroup
where $v_{ij}$ and $w_{ij}$ are similarities respectively in high and low-dimensional space. UMAP's high-dimensional conditional probabilities $v_{i|j}$ are defined as local fuzzy simplicial set memberships
\begin{align}
	v_{i|j} = \exp\left[\left(-d(x_i,x_j)-\rho_i\right)/\sigma_i\right].
\end{align}
In $v_{i|j}$, a user-selected smooth nearest neighbours distance $d(x_i,x_j)$ is defined (only Euclidean distance is used in this work), $\rho_i$ is the nearest neighbour distance~\cite{Dong2011EfficientKN} and $\sigma_i$ is an approximation for the $k$-nearest neighbour distance.

Like t-SNE the high dimensional similarities $v_{ij}$ are defined to be symmetric and follow
\begin{align}
	v_{ij} = (v_{j|i}-v_{i|j}) - v_{j|i}v_{i|j}.
\end{align}
As for the low-dimensional similarities $w_{ij}$ they follow
\begin{align}
	w_{ij} = \left(1+a||y_i-y_j||^{2b}\right)^{-1},
\end{align}
where $a$ and $b$ are user-defined parameters found through a fitting algorithm. If $a$ and $b$ are 1, we have the t-student function of t-SNE.

\subsubsection{Isometric Mapping (Isomap)}
Isometric mapping finds the nearest neighbours of every sample and creates a graph representation where every point is connected to its neighbour~\cite{tenenbaumGlobalGeometricFramework2000c}. The algorithm attempts to compute the shortest distance between every connected point. Finally, a multidimensional scaling step computes a low-dimensional graph representation.

Since this technique does not offer significant advantages (see Sec.~\ref{sec:results}) it is not discussed in further details.

\subsection{Predictive methods}
Predictive methods need to be trained using data, once trained these methods offer a transformation that can be used to label new signals. This generally translates into fast computation but requires an initialization step to train the model.

\subsubsection{Principal Component Analysis (PCA)}
Principal component analysis is a linear method previously used for TES and superconducting nanowire single-photon detector (SNSPD) signal classification~\cite{humphreys_tomography_2015,schapeler_how_2023}. For a data matrix $\bm{X}$, PCA transforms $\bm{X}$ to a new coordinate system to minimize the total distance between the samples and the principal components (columns of $\bm{W}$). By minimizing this distance, the variance of the projected points is maximized~\cite{jolliffeMathematicalStatisticalProperties2002b}. For a data matrix $\bm{X}$ and a principal component matrix $\bm{W}\in \mathbb{R}^{u \times r}$, the matrix multiplication
\begin{align}\label{eq:PCA}
	\bm{Y}=\bm{X} \bm{W},
\end{align}
transforms every signal into a low-dimensional representation $\bm{Y}\in \mathbb{R}^{u\times r}$ of size $r$ equal to the number of principal components considered. It can be shown that optimal vectors of $\bm{W}$ are given by the singular value decomposition (SVD) of the covariance matrix $\bm{X}^{\top}\bm{X}$. This is further simplified to SVD elements of $\bm{X}$ where $\bm{W}$ is taken directly from $\bm{X}=\bm{U} \bm{\Sigma} \bm{W}^{\top}$. In this decomposition, $\bm{U}$ and $\bm{V}$ are orthogonal and $\bm{\Sigma}$ is a rectangular diagonal matrix. Once $\bm{W}$ is defined, prediction is done by replacing $\bm{X}$ by new data $\bm{X}_{\text{pred}}$ in equation \eqref{eq:PCA}.

\subsubsection{Kernel Principal Component Analysis (Kernel-PCA)}
Kernel principal component analysis uses a mapping to project data onto a feature space of size $Q$ (typically $Q\gg t$) where the data has the potential of being linearly separable~\cite{Scholkopf1997KernelPC}.
It can be shown that the projection of the data points inside the feature map $\phi(x)$ onto the principal components in the feature space can be computed without explicitly computing the mapping $\phi(x)$. This is done through the introduction of a kernel function that follows some restrictions in its construction~\cite{hofmannKernelMethodsMachine2008}.

We benchmark a Polynomial (Poly), Radial Basis Function (RBF), Sigmoid and Cosine kernel defined as:
\begin{align}
	\text{Poly}:&\quad k(x_n,x_m) = (\gamma x_n^{\top}x_m + c)^d,\\
	\text{RBF}:&\quad k(x_n,x_m) = \exp\left(-\gamma||x_n-x_m||^2\right),\\
	\text{Sigmoid}:&\quad k(x_n,x_m) = \tanh(\gamma x_n^{\top}x_m+c),\\
	\text{Cosine}:&\quad k(x_n,x_m) = (x_n x_m^{\top})(||x_n||\,||x_m||)^{-1}.
\end{align}

\subsubsection{Non-Negative Matrix Factorization (NMF)}
Non-negative matrix factorization is an iterative process that attempts to find a decomposition without negative elements to minimize some objective function. The method gives an approximate decomposition of the data matrix $\bm{X}$ described by
\begin{align}
	\bm{X}\approx \bm{Y} \bm{H},
\end{align}
where $\bm{Y}$ represents the transformed data matrix and $\bm{H}$ the transformation matrix, which are both smaller matrices than $X$. The general process behind NMF offers a framework to compute adequate decompositions for specific applications. In other words, the loss function is chosen given an application. In this paper, we use a loss defined as
\begin{align}
	L(\bm{X},\bm{Y},\bm{H}) = ||\bm{X}-\bm{Y} \bm{H}||_{\text{Frob}}^2.
\end{align}
The Frobenius norm is a matrix norm defined for a matrix $\bm{A}$ with elements $a_{ij}$ as $||\bm{A}||_{\text{Frob}}=(\sum_{ij}  |a_{ij}|^2)^{1/2}$.
It can be shown that the optimization of the Frobenius norm is equivalent to the maximum likelihood estimate of $\bm{X}$ without Gaussian noise~\cite{nijs_mathematical_2021}. Additionally, we test NMF optimization using the $\mathrm{KL}$ divergence, where the loss function becomes
\begin{align}
	L(\bm{X},\bm{Y},\bm{H}) = \mathrm{KL}(\bm{X} \| \bm{Y}\bm{H}).
\end{align}
Similarly to the Frobenius norm, the use of the $\mathrm{KL}$ divergence is equivalent to the maximum likelihood estimate of $\bm{X}$ without Poissonian noise~\cite{nijs_mathematical_2021}.

To make a prediction using NMF, a new approximate decomposition is optimized based on a close-to-optimal initial guess defined in the training step.

\subsubsection{{Neural Networks}}
Neural networks have the potential to reproduce a wide variety of operations in a numerical structure that can be used efficiently to process large amounts of data. To quickly apply UMAP and t-SNE on new data, we use parametric implementations of these methods using neural networks. The main principle behind these parametric implementations is to constrain the embedding to transformations done through a neural network. In other words, a neural network is trained to optimize the $\mathrm{KL}$ divergence in the case of t-SNE and the cross-entropy in UMAP. By applying this constraint during training, we create a neural network that considers local structures and behaves similarly to t-SNE and UMAP. At this stage, new data can be embedded at an efficiency restricted by the complexity of the neural network architecture.

For both parametric implementations of t-SNE and UMAP, we use a simple feed-forward neural network defined as a series of blocks containing linear layers with ReLU activation functions followed by a batch normalization step. The results presented in this work use a neural network with 4 blocks, where each linear layer contains 300 inputs and outputs. While more complex architectures could be used for this task, we find that even an elementary neural network can achieve this task accurately, resulting in fast data transformation adequate for real-time processing.

We note that we use the same neural network to predict data for both close-to-uniform and close-to-geometric cases. This is done to train the neural network on a balanced dataset, and in this process we guarantee that the neural network is never trained on test data. Using different distributions for the training step is not advantageous to parametric methods, since the close-to-uniform dataset contains fewer samples than in the close-to-geometric case.

For more details about the implementation, the source code for parametric algorithms is available on the public repository provided in Ref.~\cite{repo}.

\subsection{Clustering}\label{sec:clustering}
Clustering refers to identifying groups of similar samples inside a latent space. For this task we use a Gaussian mixture model, given a user-defined number of clusters, this method finds the parameters of a mixture of Gaussians to describe the sample's distribution.

The choice is highly inspired by a similar model previously used in the tomography of TESs in combination with PCA~\cite{humphreys_tomography_2015}. Mixture models offer a statistical interpretation of latent spaces convenient for metrology and performance evaluation (confidence metric in Sec.~\ref{sec:conf}).

The mixture model gives a continuous probability density function for the position $s$ of samples given optimal parameters $\theta=\{(\omega_k, \mu_k$, $\Sigma_k):k=1,\cdots,K\}$. In the model, every cluster $k$ is weighted by a value $\omega_k$ (where $\sum_{k=1}^K \omega_k = 1$), and modelled by a Gaussian with mean $\mu_k$ and covariance matrices $\Sigma_k$. The individual Gaussians $\mathcal{N}$ give the cluster probability density function and the probability of observing samples in position $s$ given parameters $\theta$ are defined by
\begin{align}\label{eq:mixture}
	p(s|\theta) = \sum_{k=1}^K \omega_k \mathcal{N}(s|\mu_k , \Sigma_k).
\end{align}
The probability density function is found through an expectation maximization algorithm (EM algorithm) that attempts to find the maximum likelihood estimate of samples following a likelihood of  
\begin{align}\label{eq:likelihood}
	\mathcal{L}(\theta) = \prod_{i=1}^p \sum_{k=1}^K \omega_k \mathcal{N}(s_i|\mu_k , \Sigma_k).
\end{align}
Numerically it is more convenient to express this problem in terms of the log-likelihood given by
\begin{align}
	\ell(\theta) = \log(\mathcal{L}(\theta)) = \sum_{i=1}^p \log\left(\sum_{k=1}^K \omega_k \mathcal{N}(s_i|\mu_k , \Sigma_k)\right),
\end{align}
where the problem can be computed in terms of sum instead of products.

After computing the Gaussian mixture, photon numbers are assigned based on the average area of the samples inside each cluster. The mean area provides a reference for the cluster ordering, since a signal's area grows with energy and, consequently, the number of photons. This task is unambiguous because the average areas are well separated, unlike the individual samples.

\subsection{Number of Clusters}
\begin{figure}[!t]
	\centering
	\includegraphics[width=.45\textwidth]{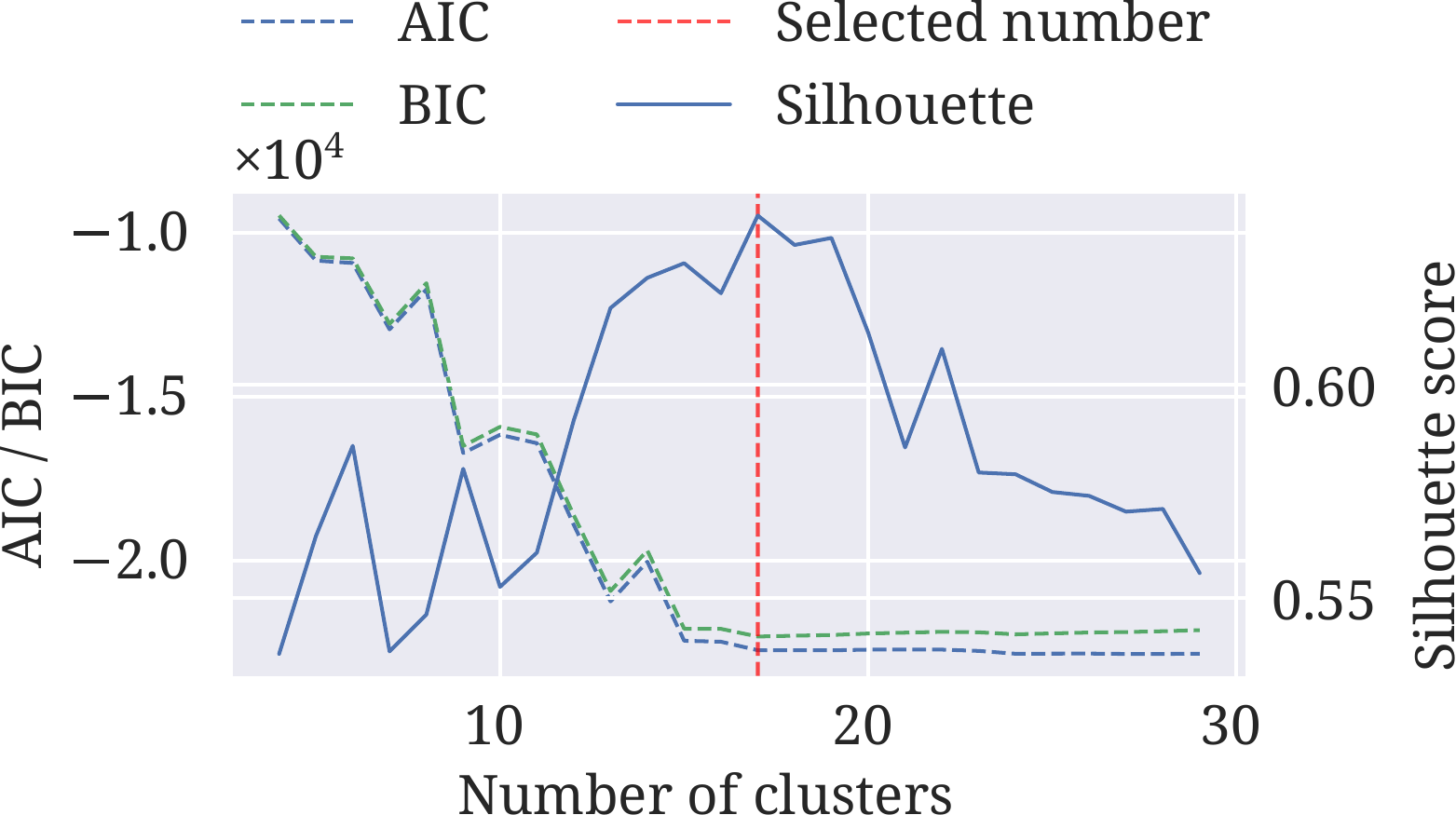}
	\caption{Akaike and Bayesian information criterion (AIC and BIC) as a function of the number of clusters in the context of a maximum value latent space on the Synthetic Uniform dataset. The point of diminishing returns gives an estimate for the number of clusters in the latent space. The Silhouette score is also plotted, where the maximum value gives an estimate for the number of clusters.}
	\label{fig:information}
\end{figure}

The Gaussian mixture model offers different advantages for quality assessment but cannot directly determine the number of clusters in a latent space. The problem is solved using an elbow method considering the Akaike information criterion ($\mathrm{AIC}$)
\begin{align}
	\mathrm{AIC} = 2K - 2\ln(\mathcal{L}(\theta))\label{eq:AIC},
\end{align}
or the Bayesian information criterion ($\mathrm{BIC}$)
\begin{align}
	\mathrm{BIC}  = K\ln(u) - 2\ln(\mathcal{L}(\theta))\label{eq:BIC}.
\end{align}
The criteria assign a score given a number of clusters $K$, a likelihood function $\mathcal{L}(\theta)$, and a total number of data points $u$. By sweeping the number of clusters used in some models, these criteria give a way to find a balance between the number of clusters and the likelihood. In our case, the likelihood of the Gaussian mixture model is used to evaluate the information scores. The general idea of these criteria is to negatively score the number of clusters, considering it is always possible to overfit the data with more clusters. In other words, a model with more clusters can always achieve a higher or equal likelihood than a model with fewer clusters. The point of diminishing return is given by the ``elbow'' of the $\mathrm{AIC}$ and $\mathrm{BIC}$ when evaluating the criteria as a function of the number of clusters, see Fig.~\ref{fig:information}. After this point, the additional clusters mostly overfit the data.
The maximum Silhouette score (see Fig.~\ref{fig:information}) is also used along the information criteria to evaluate the number of clusters~\cite{rousseeuw_silhouettes_1987}. Since similar results are found with this method, the details are not described here.

\subsection{Quality Assessment}
Assessing the performance of dimensionality reduction techniques in an unsupervised setting is difficult since the ground truth is unknown. To tackle this task, we quantify cluster separation. To improve the performance evaluation it is also important to understand that the problem is not completely unsupervised considering photon sources used to generate samples follow known distributions. We include this knowledge of photon-number distributions as an additional validation to cluster separation evaluation in the confidence metric (Sec.~\ref{sec:conf}).

\subsubsection{Confidence}\label{sec:conf}
We consider the probability density of photon events can be approximated from the sample's distribution in the latent space following the Gaussian mixture model. Following previous work~\cite{humphreys_tomography_2015}, the confidence $C_n$ is used as a performance metric for the resolution of photon numbers in a latent space, following,
\begin{align}\label{eq:conf}
C_n = \int_{-\infty}^{\infty} \frac{p(s|n)^2 P(n)}{\sum_k p(s|k) P(k)} \mathrm{d} s.
\end{align}
In this equation, $p(s|n)$ is the probability density of observing a sample in position $s$ in the latent space given it is labelled as $n$ photons. Additionally, $P(n)$ is the probability of assigning a photon number $n$. In this model, we consider that the true clusters follow a Gaussian structure inside the latent space.

The confidence represents the probability of correctly labelling a sample in a given cluster in the mixture model. We note that equation~\ref{eq:conf} describes the confidence for a one-dimensional space but can be generalized to an arbitrarily high-dimensional latent space.

In practice $p(s|n)$ can be measured using a trusted source of $n$ photons (i.e. using detector tomography), in which case $C_n$ is equal to the probability of the detector measuring and assigning the correct number of photons~\cite{humphreys_tomography_2015}.

It is important to mention that the distances in the latent space do not necessarily have a physical meaning. The separation must only be interpreted as our capacity to distinguish clusters, and the confidence translates this concept into a probabilistic framework.

\subsection{Datasets}\label{sec:dataset}
\begin{figure*}[t]
	\centering
	\begin{subfigure}[t]{0.45\textwidth}
    	\centering
    	\includegraphics[width=\textwidth]{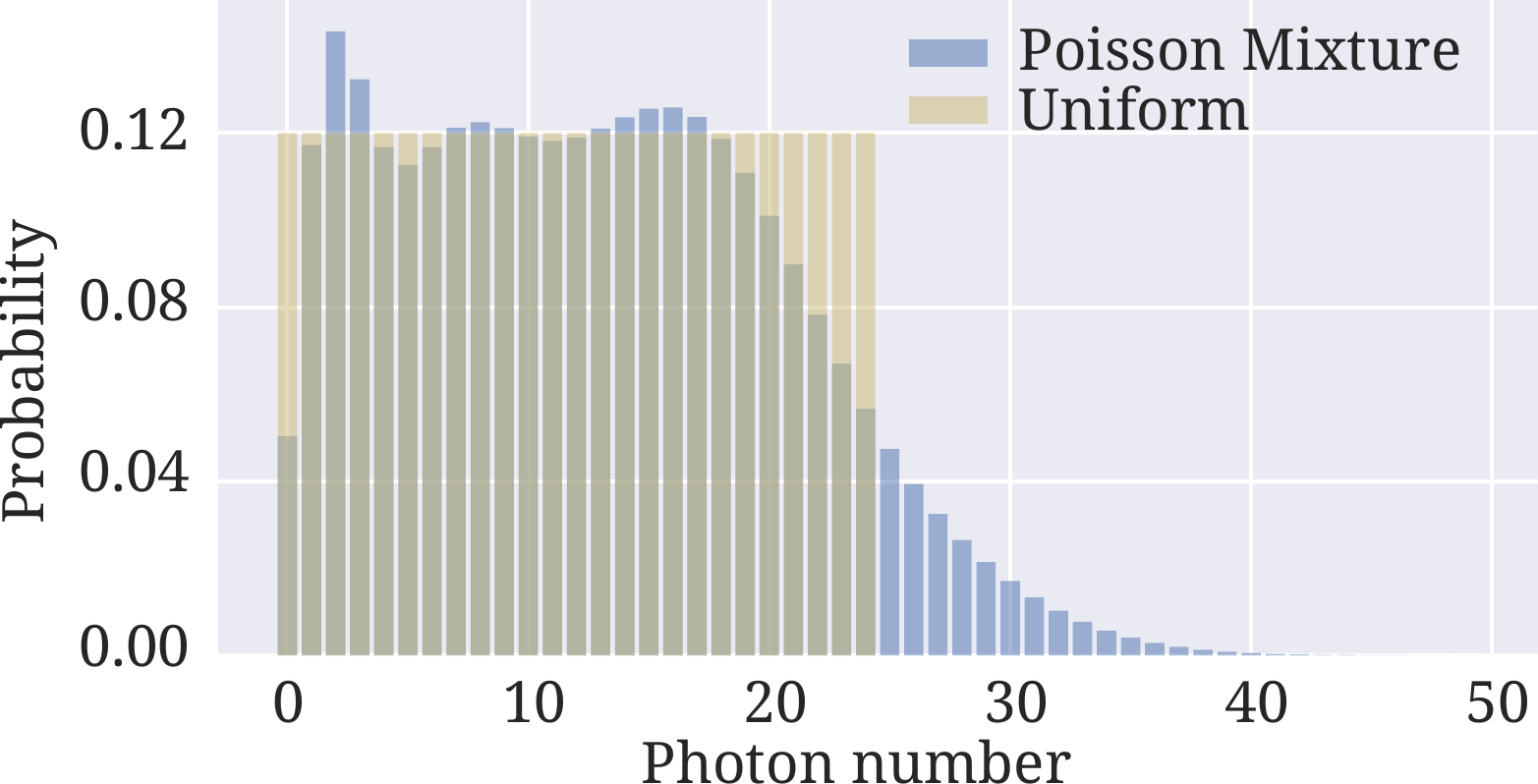}
    	\caption{\textbf{Synthetic Uniform : } Composed of $u=30\,550$ train and test samples of size $t=350$, generated using an attenuated 1550 nm coherent source~\cite{gerrits_extending_2012,Gerrits2024-ap}.}
    	\label{fig:UniformDist}
	\end{subfigure}
	\hfill
	\begin{subfigure}[t]{0.45\textwidth}
    	\centering
    	\includegraphics[width=\textwidth]{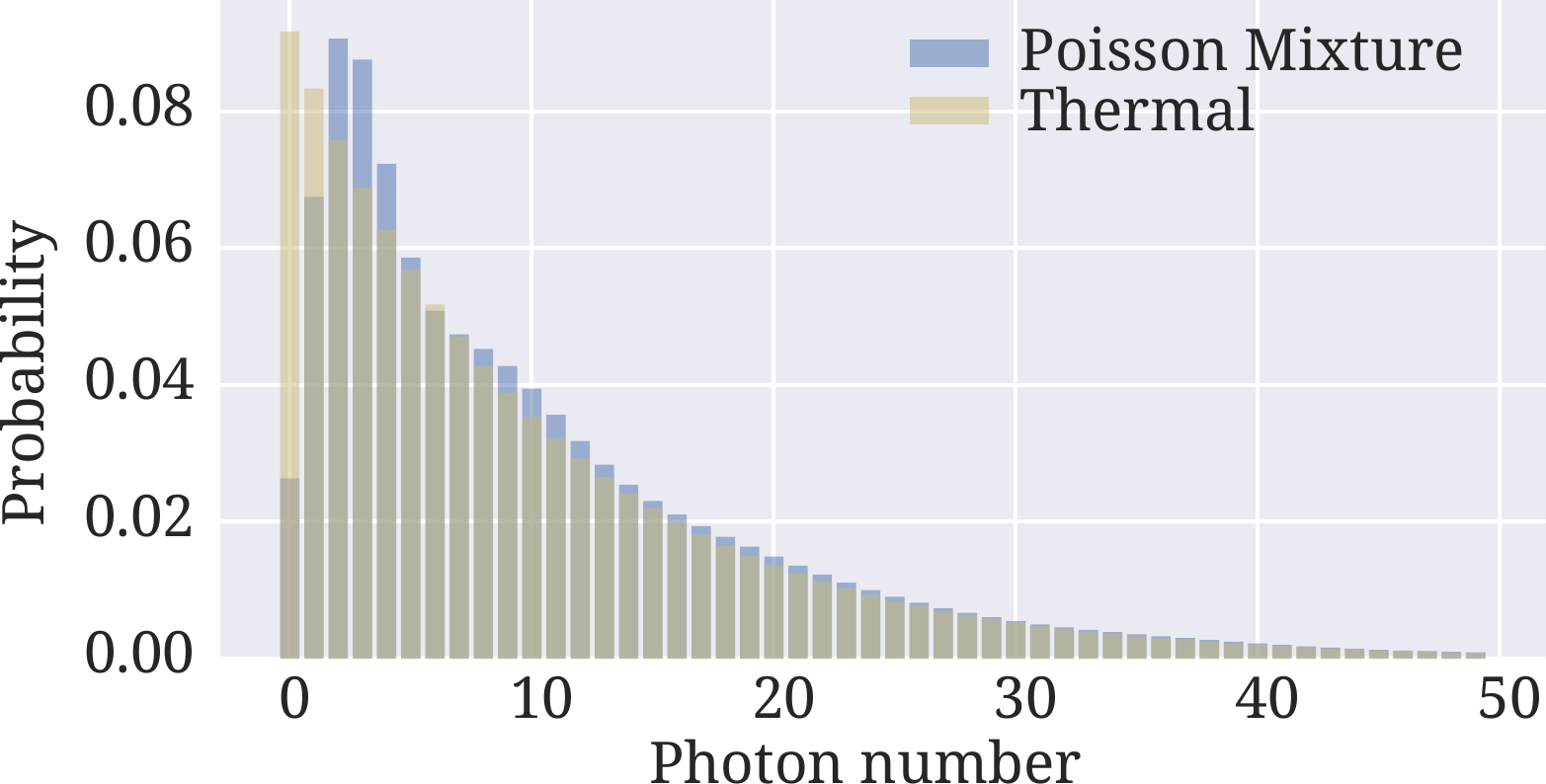}
    	\caption{\textbf{Synthetic Geometric : } Composed of $u=57\,020$ train and test samples of size $t=350$, generated using an attenuated 1550 nm coherent source~\cite{gerrits_extending_2012,Gerrits2024-ap}.}
    	\label{fig:ThermalDist}
	\end{subfigure}
	\hfill
	\begin{subfigure}[t]{0.45\textwidth}
    	\centering
    	\includegraphics[width=\textwidth]{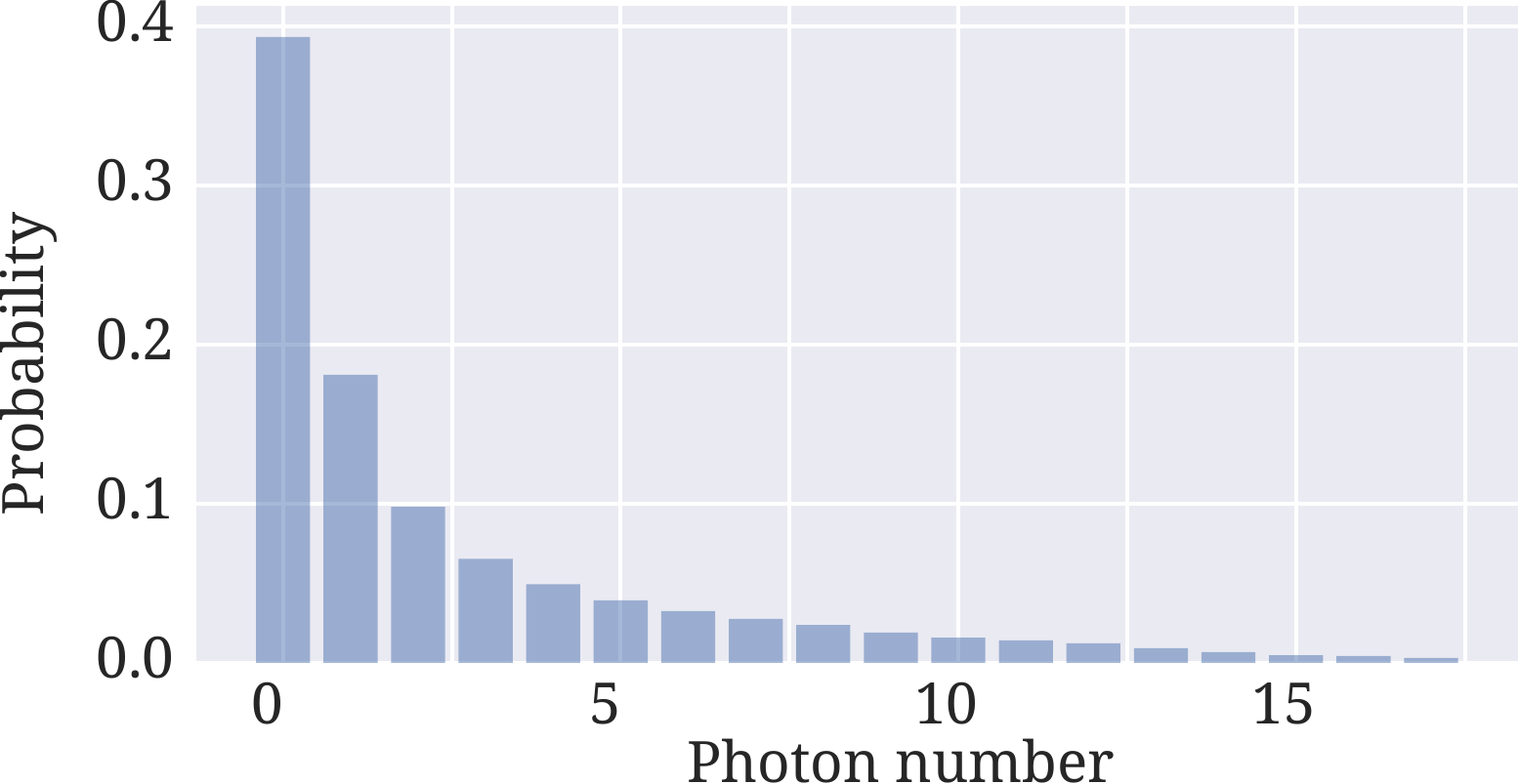}
    	\caption{\textbf{Synthetic Large : } Composed of $u=550\,000$ train and test samples of size $t=100$, generated using a 1530nm coherent source~\cite{Dalbec-Constant2024-ap}.}
    	\label{fig:NRCDist}
	\end{subfigure}
	\hfill
	\begin{subfigure}[t]{0.45\textwidth}
    \centering
		\makebox[0.45\textwidth]{
			\includegraphics[width=0.8\textwidth]{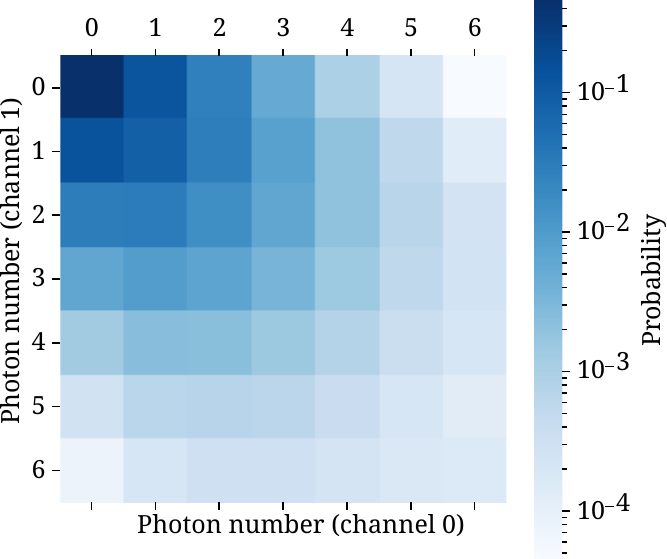} 
		}
		\caption{\textbf{Noise : } Composed of $u=1\,000\,000$ samples in both channels of size $t=50$, generated using twin beams at 1520 nm and 1580 nm. Noise at different wavelengths is also present in the data and is filtered out using UMAP in the plotted distribution (not characterized)~\cite{Dalbec-Constant2024-ap}.}
		\label{fig:NoiseDist}
	\end{subfigure}
	\caption{Compilation of all datasets used in this research. For cases where the photon-number distribution is engineered to resemble a goal distribution, the \textbf{blue bars} represent the expected photon-number distribution for a mixture of Poisson distribution and the \textbf{yellow bars} are the goal distributions used to fit the weights $w_{\langle n \rangle}$.}
	\label{fig:dataset}
\end{figure*}

Experimental data from previous work at the National Institute of Standards and Technology (NIST) is used to benchmark the different techniques in this work~\cite{gerrits_extending_2012}. The original dataset was generated by progressively attenuating a coherent source from 29dB to 7dB, leading to 24 datasets each containing $u=20\,480$ signals and $t=8\,192$ time steps. This results in datasets that each have Poisson photon-number distributions and mean photon number $\braket{n_1}=2.26$ to $\braket{n_{24}}=7.08\times 10^6$. These values were independently measured using a calibrated photodetector.

Instead of directly using these distributions, we construct two synthetic datasets (made of real traces) that follow a close-to-uniform and close-to-geometric distribution $P(n)$. These datasets are labelled as Synthetic Uniform and Synthetic Geometric in Fig.~\ref{fig:dataset}. Furthermore, for all datasets, a training and testing set were generated by taking respectively the first and second half of the files for every attenuation. This simulates a calibration and measurement stage but effectively does not change the results. Considering randomly selecting a portion of the samples in each experiment is equivalent to varying the weight $w_{{\langle n \rangle}}$ of a given Poisson distribution $P_{{\langle n \rangle}}(n)$ inside a mixture of Poisson distributions. The total expected distribution $P(n)$ can be described by
\begin{align}\label{eq:photonDistribution}
	P(n) = \frac{1}{\xi} \sum_{\langle n \rangle\in \bar{N}} w_{\langle n \rangle} P_{\langle n \rangle}(n),
\end{align}
with
\begin{align}
	\xi = \sum_{\langle n \rangle \in \bar{N}} w_{\langle n \rangle},
\end{align}
and where $\bar{N}$ is the set of available mean photon numbers $\langle n \rangle$. With this construction, the expected photon-number distribution is a mixture of Poisson distributions shown in Fig.~\ref{fig:dataset}. The choice of a uniform distribution is motivated by the desire to make the labelling task difficult by maximizing the distribution's entropy. In other words, for every sample in a perfectly uniform distribution, the method would have equal chances of guessing every class. The choice of testing a geometric distribution comes from the desire to precisely measure thermal optical sources that follow a geometric photon-number distribution. Also, distributions with a long tail can be difficult to process for certain methods since fewer examples are present in some classes (imbalanced dataset).

Additionally, these expected distributions are used as $P(n)$ in the computation of the confidence. The predictive methods are trained with the training set, and the analysis of performance metrics is done by feeding the test set to the trained methods. In the case of non-predictive and basic feature methods, the test set is directly used. The training and test datasets contain a total of $u=30\,550$ traces of size $t=350$ (first $350$ values of the $8192$ available time steps). We note that most of the weights $w_{\braket{n}}$ are set to zero because of the number of available Poisson distributions in the desired photon number range is small, making the synthetic distribution not perfectly uniform (see top row in Fig.~\ref{fig:dataset}).

To validate a hypothesis discussed in Sec.~\ref{sec:limitParam} we also use a larger dataset named Synthetic Large that was created using signals generated by TESs at the National Research Council Canada (NRC) in Ottawa~\cite{Dalbec-Constant2024-ap}. The data was generated by tuning the attenuation of a 1530 nm laser and measuring $u=100 \,000$ signals for each of these coherent sources.

Finally, we also make use of a dataset labelled Noise, in Sec.~\ref{sec:outlier}, the light is generated by an integrated optical parametric oscillator (OPO) pumped below threshold using a pulsed-carved continuous wave laser, as in Ref.~\cite{vaidyaBroadbandQuadraturesqueezedVacuum2020b}. The OPO generated correlated 1520 nm and 1580 nm photons following twin beams photon statistics. In addition, noise photons from the pump leaked into both modes due to imperfect pulse carving and filtering. These photons were generated at random times relative to the signal photons, which reduced the photon correlation strength between the two modes. Tracing over one of these modes, we expect quasi-thermal photon statistics. For this dataset, $u=1\,000\,000$ TES signals were recorded from each detector channel. All datasets are summarized in Fig.~\ref{fig:dataset}.

\section{Results}\label{sec:results}
\subsection{Validation}
Before looking at performance metrics, a sanity check is done to validate the basic characteristics of the Synthetic Uniform dataset. This is done for the data from the different coherent sources (all with different mean photon numbers). Since coherent sources are used to generate the samples, a $g^{(2)}(0)$ (second-order coherence) of 1 is expected. This quantity is defined in terms of the first two moments of the photon-number distribution, as
\begin{align}
	g^{(2)}(0) = \frac{\langle n^2 \rangle - \langle n \rangle}{\langle n \rangle^2}.
\end{align}
We use the $g^{(2)}(0)$ as a validation metric both ways, by making sure the statistics of the light are correct and that the generated statistics using the numerical methods follow the physics of the system.
\begin{figure}[!t]
	\centering
	\includegraphics[width=.45\textwidth]{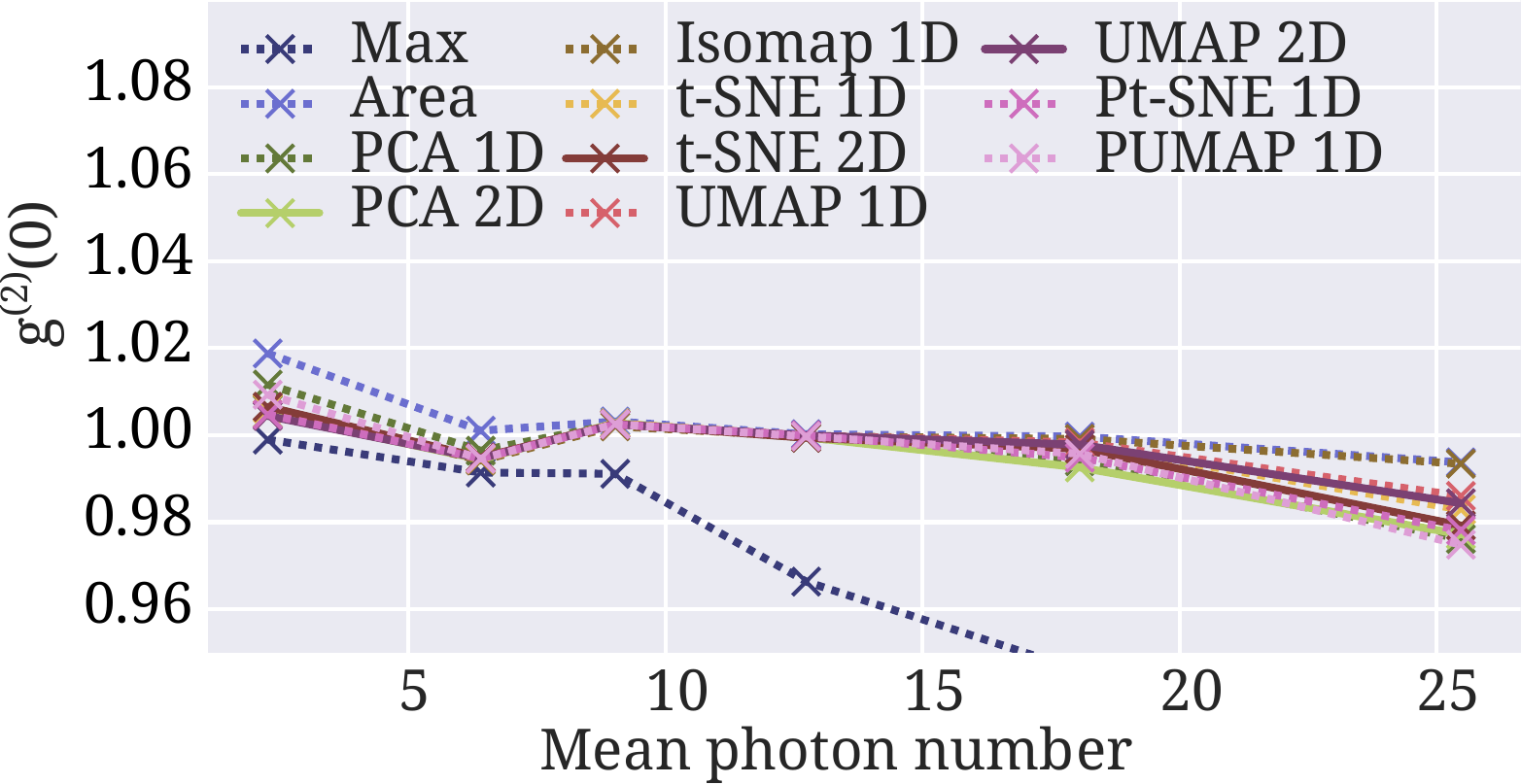}
	\caption{Computed second-order coherence for each method used on the individual datasets composing the Synthetic Uniform dataset (where markers are the mean photon number of the available coherent sources). In this figure, and the ones that follow, methods using a 1D latent space are represented by dotted lines, while those with 2D latent spaces are shown with solid lines.}
	\label{fig:g2}
\end{figure}
In Fig.~\ref{fig:g2} we can see that every method has a $g^{(2)}(0)$ close to 1 for most datasets. All methods consistently get farther from one as the mean photon number increases, the lack of resolution for high photon numbers explains this behaviour. Additionally, the number of signals associated with the high mean is limited compared to the low mean cases. The lack of resolution is especially present for the method based on the maximum value of the signals, since it cannot resolve photon numbers higher than 10 in our dataset.

\subsection{Confidence}
\begin{figure*}[!t]
	\centering
	\includegraphics[width=.95\textwidth]{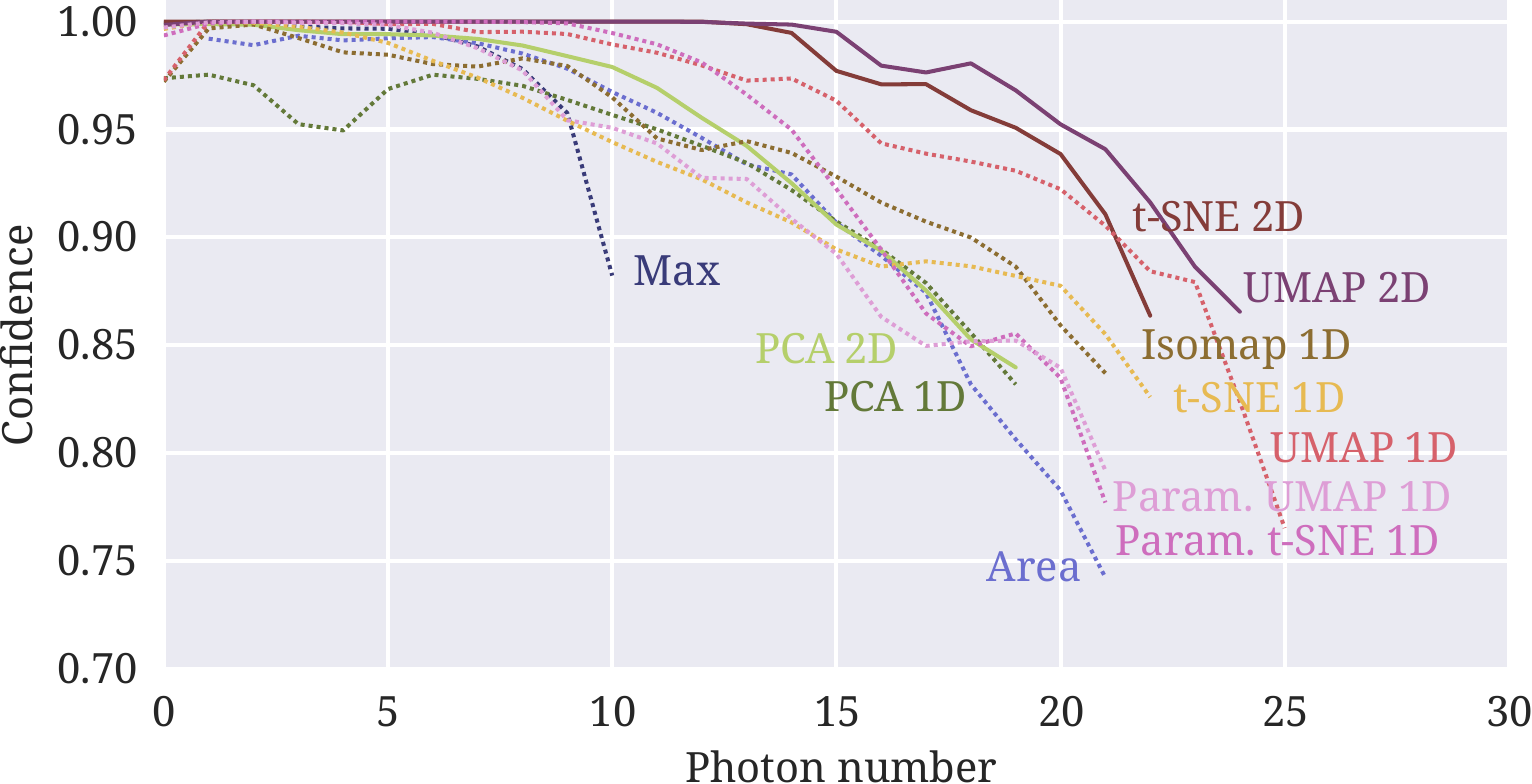}
	\caption{Confidence of photon-number clusters for the different methods using the Synthetic Uniform dataset. In this figure, and the ones that follow, methods using a 1D latent space are represented by dotted lines, while those with 2D latent spaces are shown with solid lines.}
	\label{fig:resultConf}
\end{figure*}

Considering the different dimensionality reduction techniques and following Gaussian mixture clustering, the confidence associated with every method is compiled in Fig.~\ref{fig:resultConf} for the Synthetic Uniform dataset. In this plot, the Kernel PCA techniques and NMF are not presented to facilitate readability, since they do not offer significant differences with PCA or are significantly worse.
\begin{figure}[htbp]
	\centering
	\includegraphics[width=.45\textwidth]{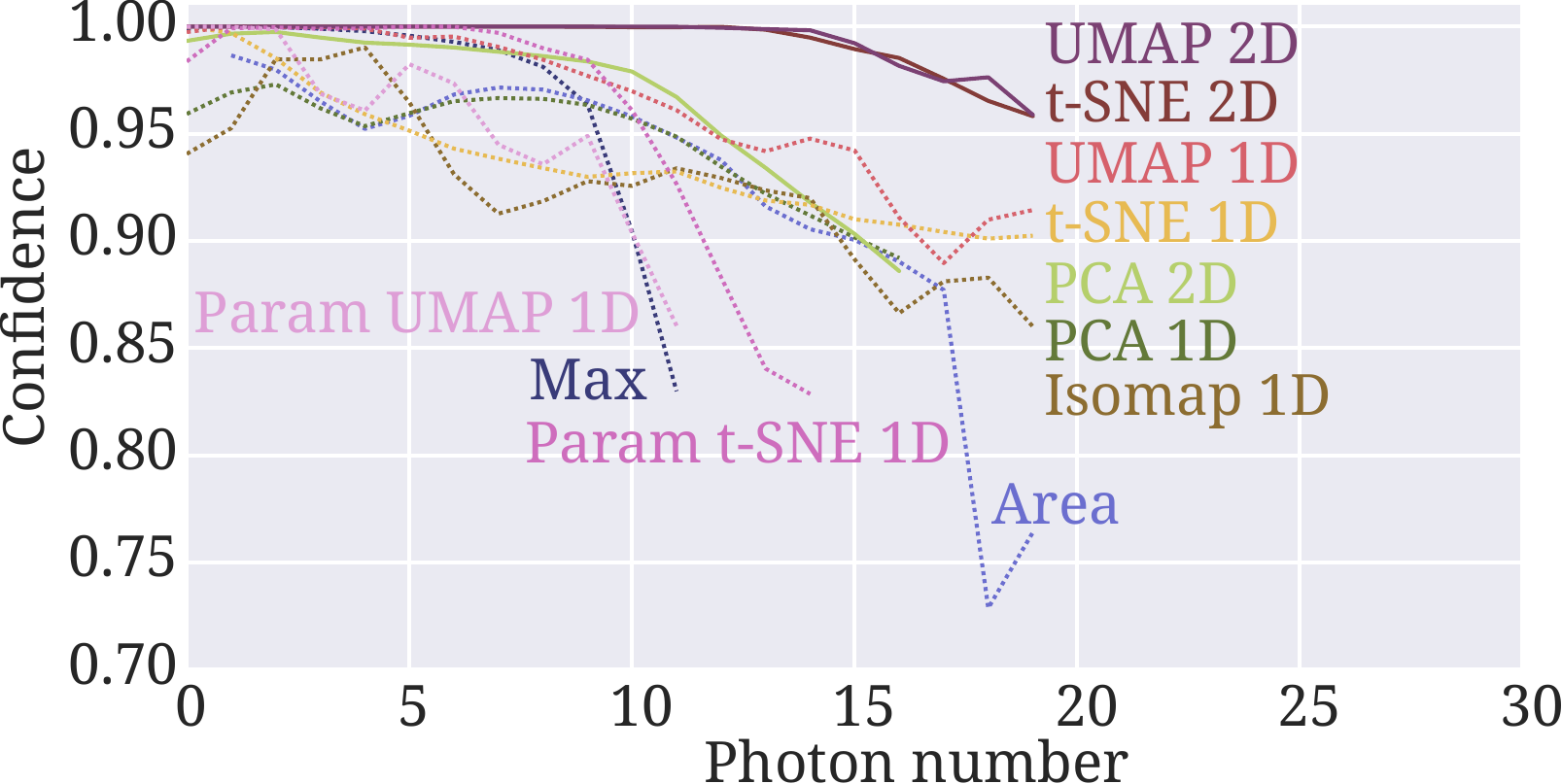}
	\caption{Confidence of photon-number clusters for the different methods using the Synthetic Geometric dataset}
	\label{fig:resultConfTherm}
\end{figure}
The number of clusters considered in the confidence plots is defined using the AIC and BIC information criteria and other considerations. First, the last cluster is always removed since it often offers an artificially high confidence considering there is no other cluster to overlap with further in the latent space. Additionally, regions associated with multiple photons described by a uniform density are ignored. This is done since regions of uniform density can be described by an arbitrarily large number of Gaussians.

We found a significant increase in performance can be achieved using nonlinear methods. In Fig.~\ref{fig:resultConf} and Fig.~\ref{fig:resultConfTherm} we show the confidence metric for the different methods considered for both the Synthetic Uniform and Synthetic Geometric datasets. We see that for both datasets previous methods like the signal's area and PCA can resolve up to 16 photons with confidence above $90\%$ while t-SNE and UMAP can resolve up to 21 with the same confidence threshold. Parametric implementations of t-SNE and UMAP did not give satisfying results for these datasets however, in Sec.~\ref{sec:limitParam} we show that these implementations can outperform PCA if the dataset is sufficiently large.

\section{Discussion}\label{sec:discussion}
\subsection{Qualitative Analysis}
\begin{figure*}[t]
	\centering
	\begin{subfigure}[b]{0.32\textwidth}
    	\centering
    	\includegraphics[width=\textwidth]{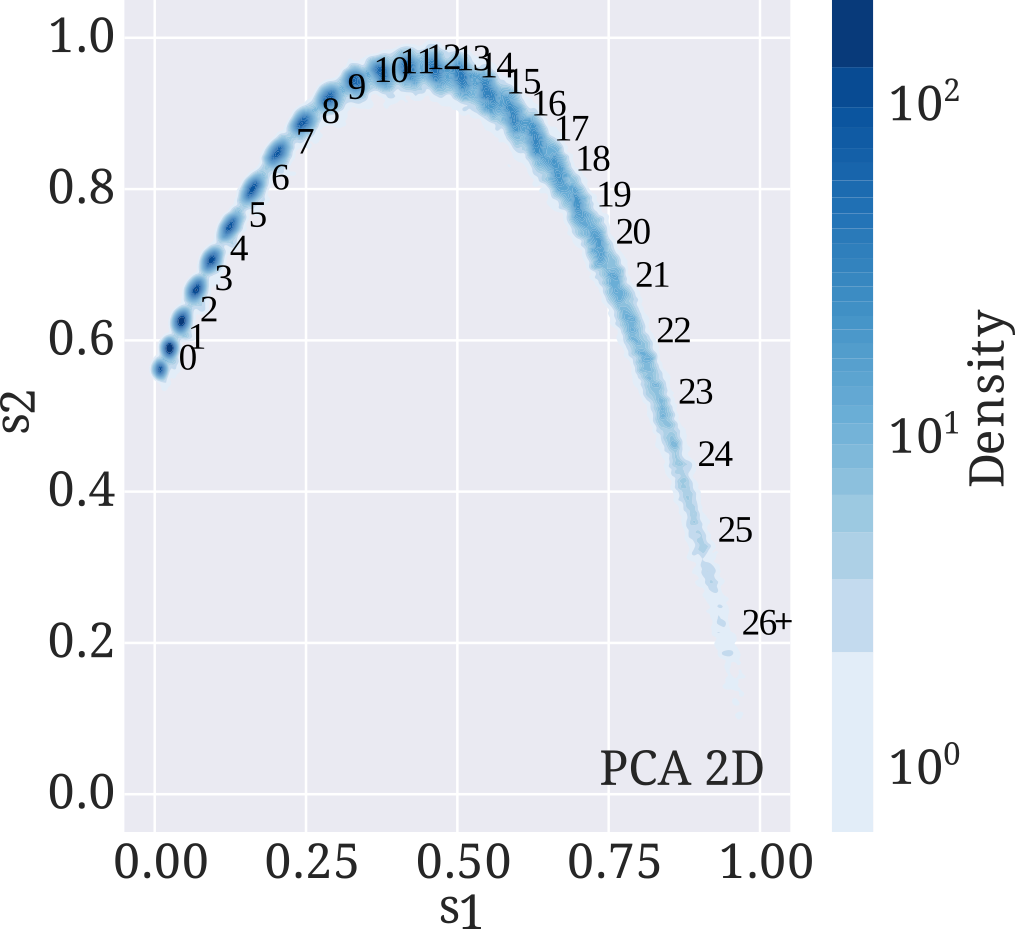}
    	\caption{}
    	\label{fig:PCA2D}
	\end{subfigure}
	\hfill
	\begin{subfigure}[b]{0.32\textwidth}
    	\centering
    	\includegraphics[width=\textwidth]{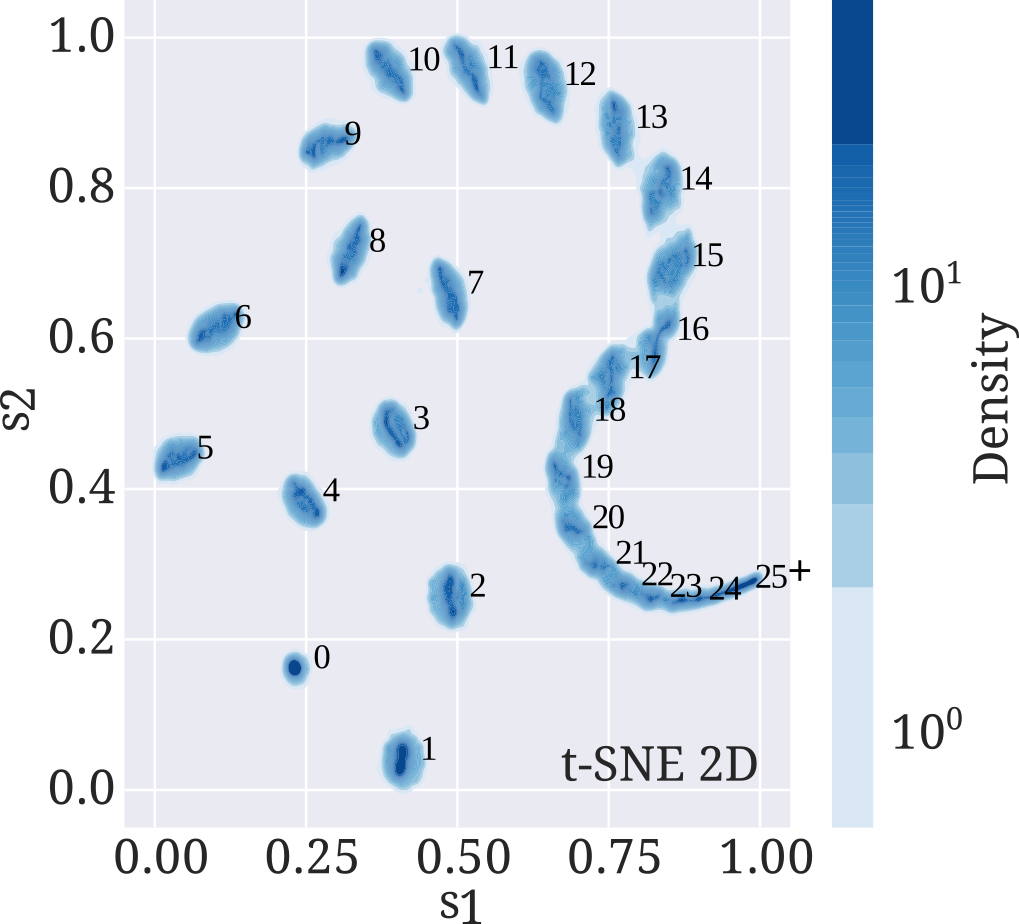}
    	\caption{}
    	\label{fig:tSNE2D}
	\end{subfigure}
	\hfill
	\begin{subfigure}[b]{0.32\textwidth}
    	\centering
    	\includegraphics[width=\textwidth]{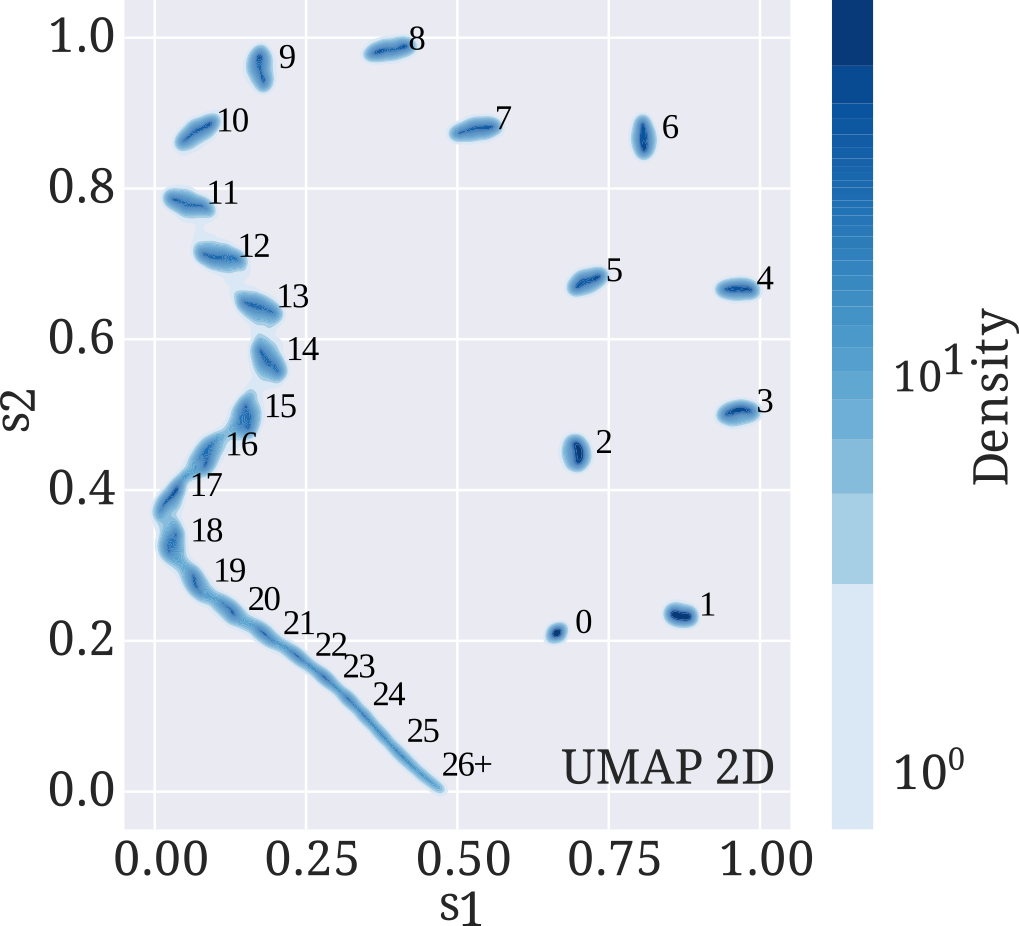}
    	\caption{}
    	\label{fig:UMAP2D}
	\end{subfigure}
	\hfill
	\begin{subfigure}[b]{0.32\textwidth}
    	\centering
    	\includegraphics[width=\textwidth]{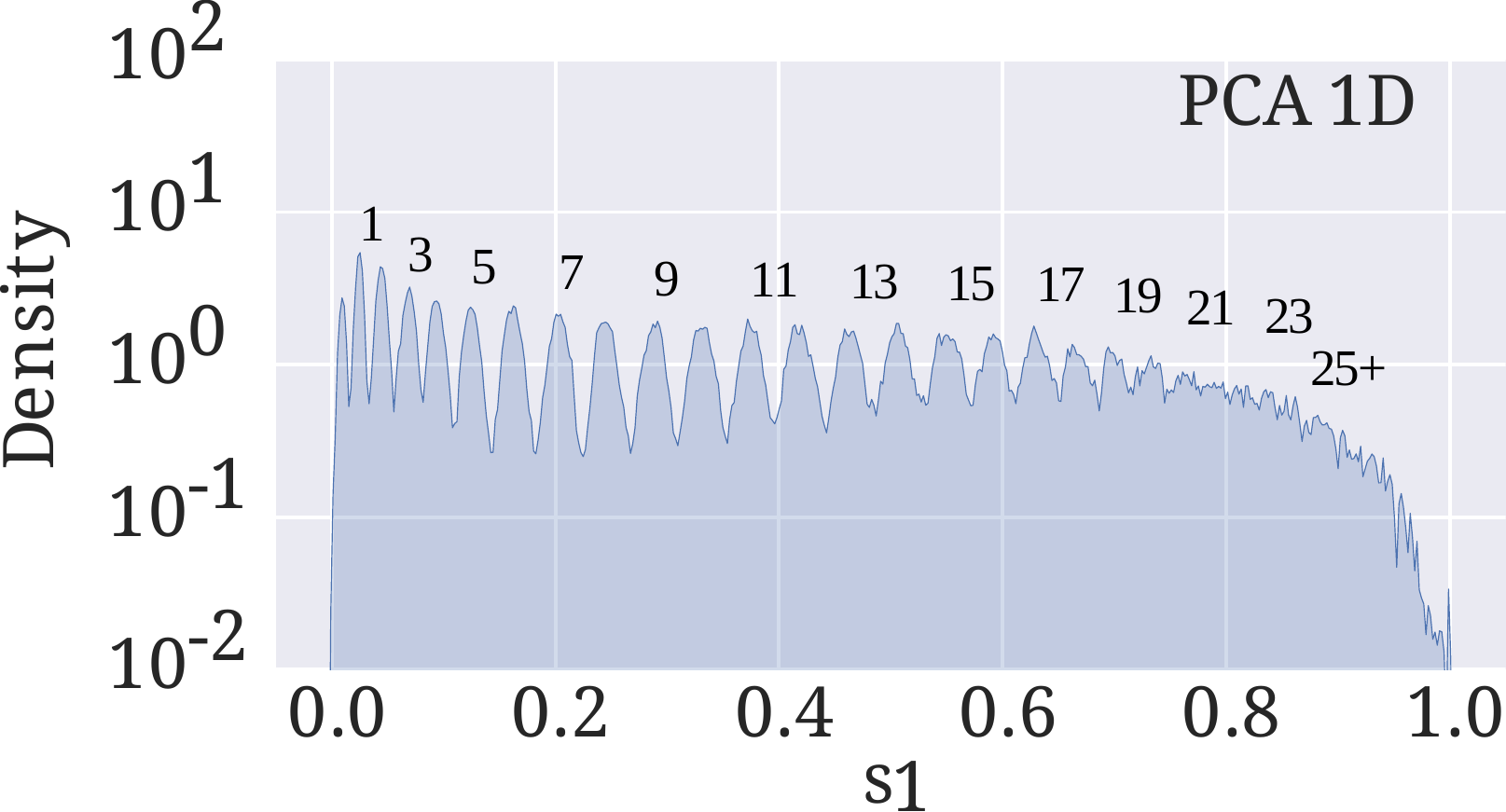}
    	\caption{}
    	\label{fig:PCA1D}
	\end{subfigure}
	\hfill
	\begin{subfigure}[b]{0.32\textwidth}
    	\centering
    	\includegraphics[width=\textwidth]{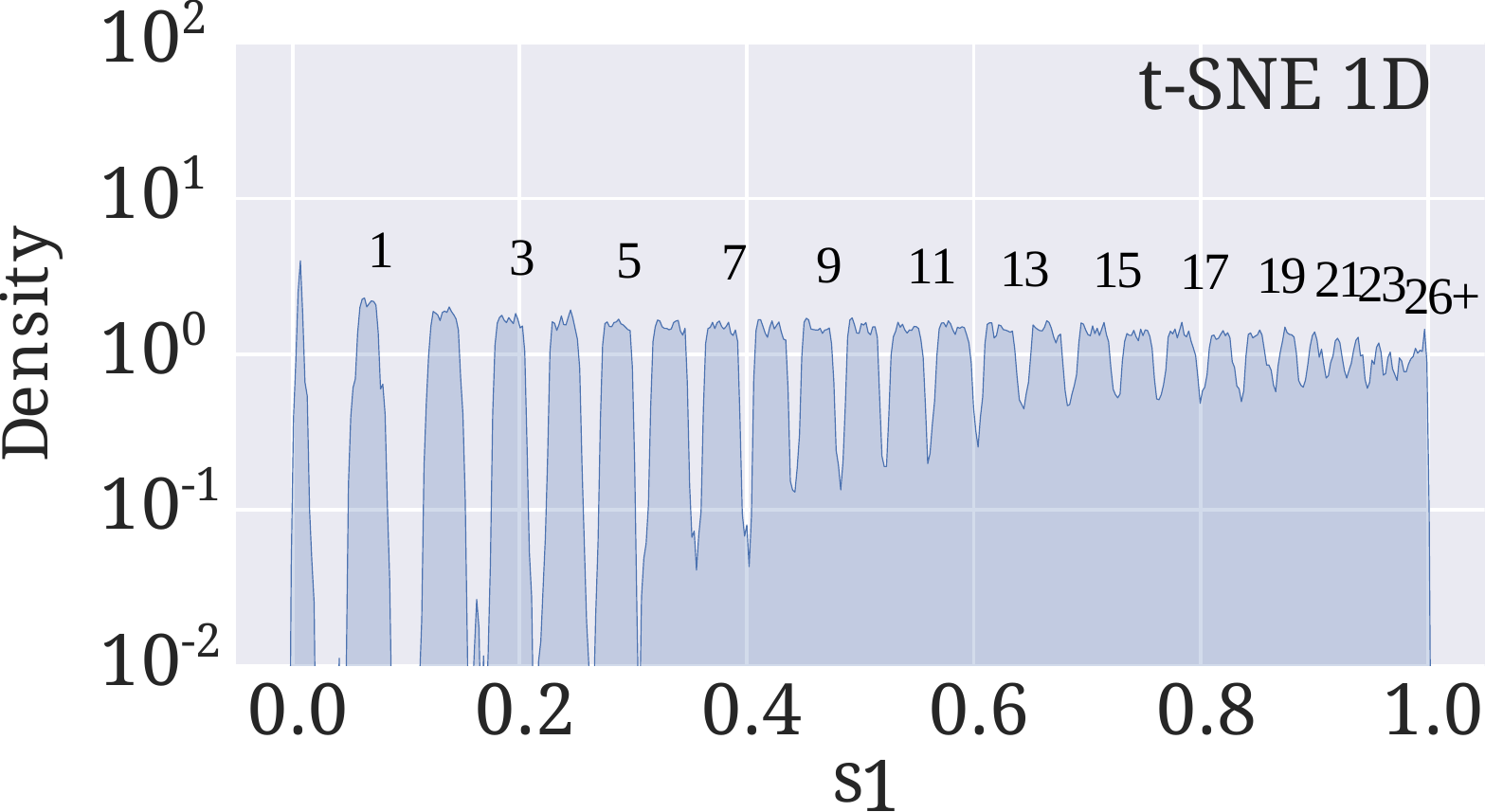}
    	\caption{}
    	\label{fig:tSNE1D}
	\end{subfigure}
	\hfill
	\begin{subfigure}[b]{0.32\textwidth}
    	\centering
    	\includegraphics[width=\textwidth]{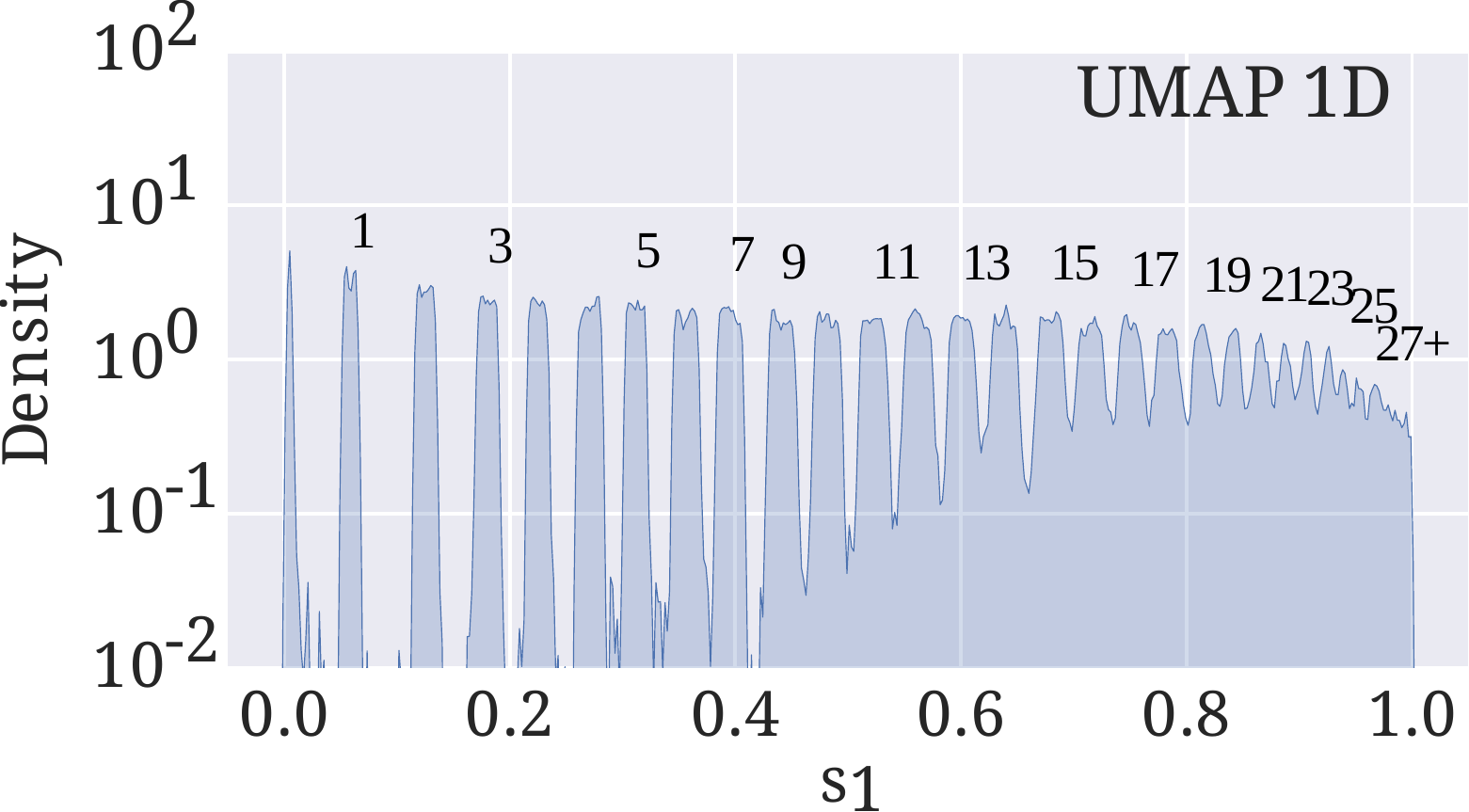}
    	\caption{}
    	\label{fig:UMAP1D}
	\end{subfigure}
	\caption{Kernel density estimation of the low dimensional embedding of TES signals generated by \textbf{(\ref{fig:PCA2D})} PCA 2D, \textbf{(\ref{fig:tSNE2D})} t-SNE 2D, \textbf{(\ref{fig:UMAP2D})} UMAP 2D, \textbf{(\ref{fig:PCA1D})} PCA 1D, \textbf{(\ref{fig:tSNE1D})} t-SNE 1D, \textbf{(\ref{fig:UMAP1D})} UMAP 1D. The clusters are labelled by sorting the clusters by the mean area of the samples inside each cluster, since the trace area is expected to increase with the photon number.}
	\label{fig:lowDim}
\end{figure*}
Through a visual analysis of the sample's distributions in latent spaces, it is possible to identify methods that show potential for unsupervised classification. In other words, methods that visually offer clear cluster separation have the potential to better perform at the classification task. To visualize the data in these different spaces, we use kernel density estimation, which involves summing a kernel function (Gaussian in this case) over all the samples to provide a smooth representation of the data distribution.

PCA is the first interesting method, since it was previously used for this task. We observe clear clusters, and the samples follow the expected arc-like structure presented in Fig.~\ref{fig:PCA2D} and observed in previous work~\cite{humphreys_tomography_2015}.

We also notice the promising separation of clusters using both t-SNE and UMAP. The sample distributions generated by these methods in two dimensions are presented in Fig.~\ref{fig:tSNE2D} and Fig.~\ref{fig:UMAP2D}.

The other methods tested in this work generate sample distributions with no special properties and, for this reason, are not further discussed. However, all methods and their results are available online~\cite{repo}.

\subsection{Limits for Parametric Implementations}\label{sec:limitParam}

We consider t-SNE and UMAP to offer some approximate upper bound on the confidence of their parametric implementation. This is justified by the fact that both methods follow the same optimization scheme. However, non-parametric methods are not limited by the set of possible transformations in the neural network architecture. We therefore hypothesize that given a large enough neural network and adequate hyperparameters, the performance of Parametric t-SNE and UMAP has the potential to resemble their non-parametric equivalent.

\begin{figure}[htbp]
	\centering
	\includegraphics[width=.45\textwidth]{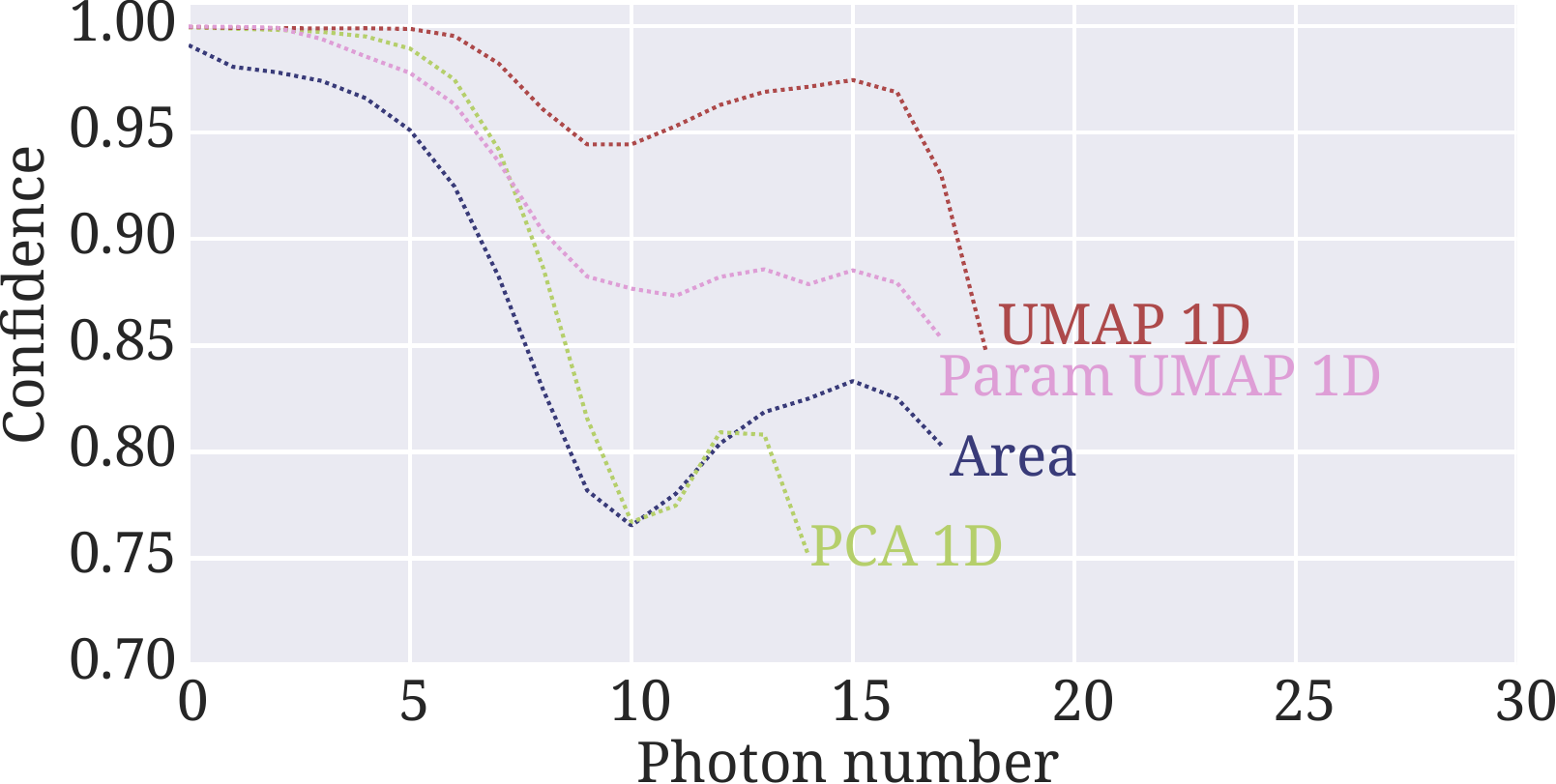}
	\caption{Confidence of Parametric UMAP compared with the non-parametric implementation and 1D PCA, for the Synthetic Large dataset taken at the National Research Council in Ottawa.}
	\label{fig:Confidence_NET}
\end{figure}

The training process to generate a network with the reported performance for the Synthetic Uniform and Geometric datasets required a fair amount of tuning to give satisfying results, which is not ideal for experimental setups. We mainly attribute this problem to the limited amount of training data, which makes it easy to overfit the model to the training data. More precisely, by learning local data structures the neural network learns less generalized features which limits its capacity to make predictions. This family of neural networks is therefore more reliant on having access to a large training dataset, since it needs examples for a wider range of fine signal features. This limits the performance capabilities demonstrated in this work, however, with a larger training set the neural networks can have prediction capabilities similar to the transformation of their non-parametric implementation. To verify this intuition, we used the Synthetic Large dataset previously mentioned in section~\ref{sec:dataset}. Using the $u=300 \,000$ signals, we trained a small feedforward neural network (5 linear layers of size 300). We present in Fig.~\ref{fig:Confidence_NET} that with sufficient data, this network offers advantageous confidence values compared to previously used techniques in one-dimension, which was not the case using less data. We limit this comparison to one-dimensional embeddings, as our focus is on techniques suitable for real-time signal processing, and the labelling task remains computationally efficient only in one dimension.

\subsection{Impact of Embedding Dimension}
The analysis of the dimensionality reduction techniques in this work assumes that the underlying true classes are associated with the photon numbers. This gives satisfying results because the traces for each photon number follow a clear pattern that different methods can easily capture. However, additional considerations are needed to solve the photon-number classification problem. First, cluster distinguishability inside the low-dimensional representations is possible because the underlying structures of photon numbers are dominant in comparison to other characteristics like noise. Additionally, the dimensionality reduction techniques are only aware of data structure at different scales and never explicitly have a grasp of the physical system. We emphasize this property since it makes the method almost completely independent of the statistics of the measured light and does not require prior knowledge of the light source. To come back to the data structures, when methods encode data in a low dimensional space they need to find a representation that describes the entire complexity of the signals. This means that noise and photon-number structures are equally preserved in the embedding. If enough noise structures exist, the method will not have enough space in a single dimension to represent this variety, and the resulting embedding can show excessive broadening of clusters. The constraint of preserving structures in the data limits the potential of finding well separated clusters in lower dimensional embeddings. This is a reason why it is easier to find an embedding with well-separated clusters in two-dimensional spaces, even if the underlying classes we wish to identify are contained in a single dimension: the photon number.

\subsection{Global vs Local Data Structures}
In unsupervised classification tasks, it is often suggested to use dimensionality reduction techniques that preserve global structures rather than local structures~\cite{umap_nodate}. This is because preserving the local structure may alter the distances and density of the data from the original space to the generated embedding. This characteristic makes it harder to guarantee that generated clusters are real or associated with the desired classes. Depending on the data, noise structures can also be grouped, creating artificial clusters. While this can be true, in the case of TES traces we argue that data does not contain electrical noise important enough to create artificial clusters. Additionally, noise from temporally uncorrelated photons is described by well-defined signal signatures. Looking at local structures gives the capacity to cluster these structures, arguably making it a positive rather than a negative feature, as we explain in the next section.

\subsection{Outlier Detection}\label{sec:outlier}
\begin{figure}[htbp]
 	\centering
 	\begin{subfigure}[b]{0.45\textwidth}
     	\centering
     	\includegraphics[width=\textwidth]{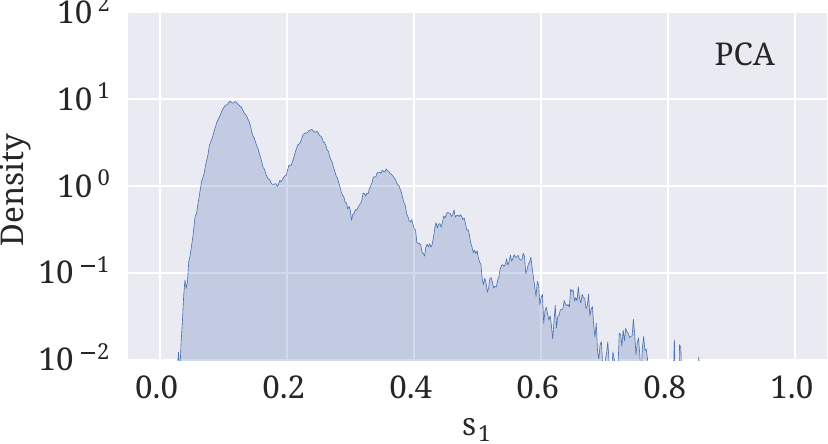}
     	\caption{Density estimation of PCA embedding using the first principal component.}
     	\label{fig:noisePCA1D}
 	\end{subfigure}
 	\hfill
 	\begin{subfigure}[b]{0.45\textwidth}
     	\centering
     	\includegraphics[width=\textwidth]{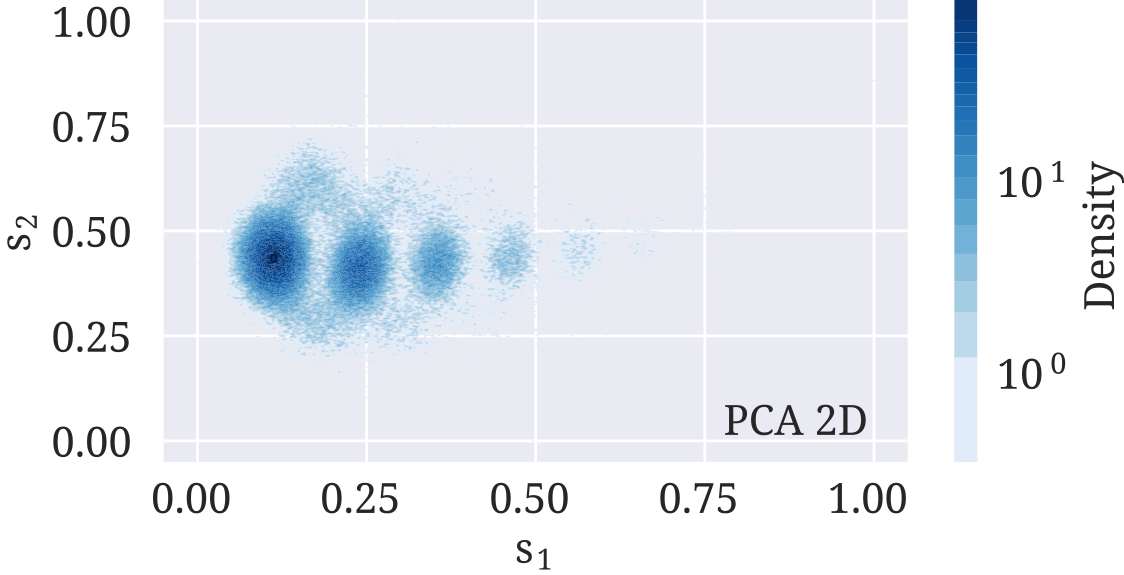}
     	\caption{Scatter plot of embedding of TES traces using PCA in two dimensions.}
     	\label{fig:noisePCA2D}
 	\end{subfigure}
	\caption{Low dimensional representation using PCA of the Noise dataset containing signals from a system with temporally uncorrelated photons.}
	\label{fig:noisePCA}
\end{figure}

A one-dimensional embedding is efficient from a computational point of view, since the clustering problem can be translated into a sorted array search. However, depending on the use case, we argue that two dimensions may offer deeper insight due to their capacity to capture a wider range of structures. For example, if temporally uncorrelated light overlaps with the light modes one seeks to analyze, then a single dimension is likely not enough space to correctly capture the photon-number statistics of the modes under analysis. Adding to what is mentioned in the previous section, the noise becomes an additional structure to represent, and effectively the proportion of information that the method can allocate to the photon-number structure is reduced. This is shown in Fig.~\ref{fig:noisePCA1D} where we use a single channel in the Noise dataset (section~\ref{sec:dataset}) and observe cluster broadening due to the presence of temporally uncorrelated photons. In this case, the two-dimensional representation becomes more useful to describe the complexity of the dataset (see Fig.~\ref{fig:noisePCA2D}).
\begin{figure*}[t]
	\centering
	\includegraphics[width=.95\linewidth]{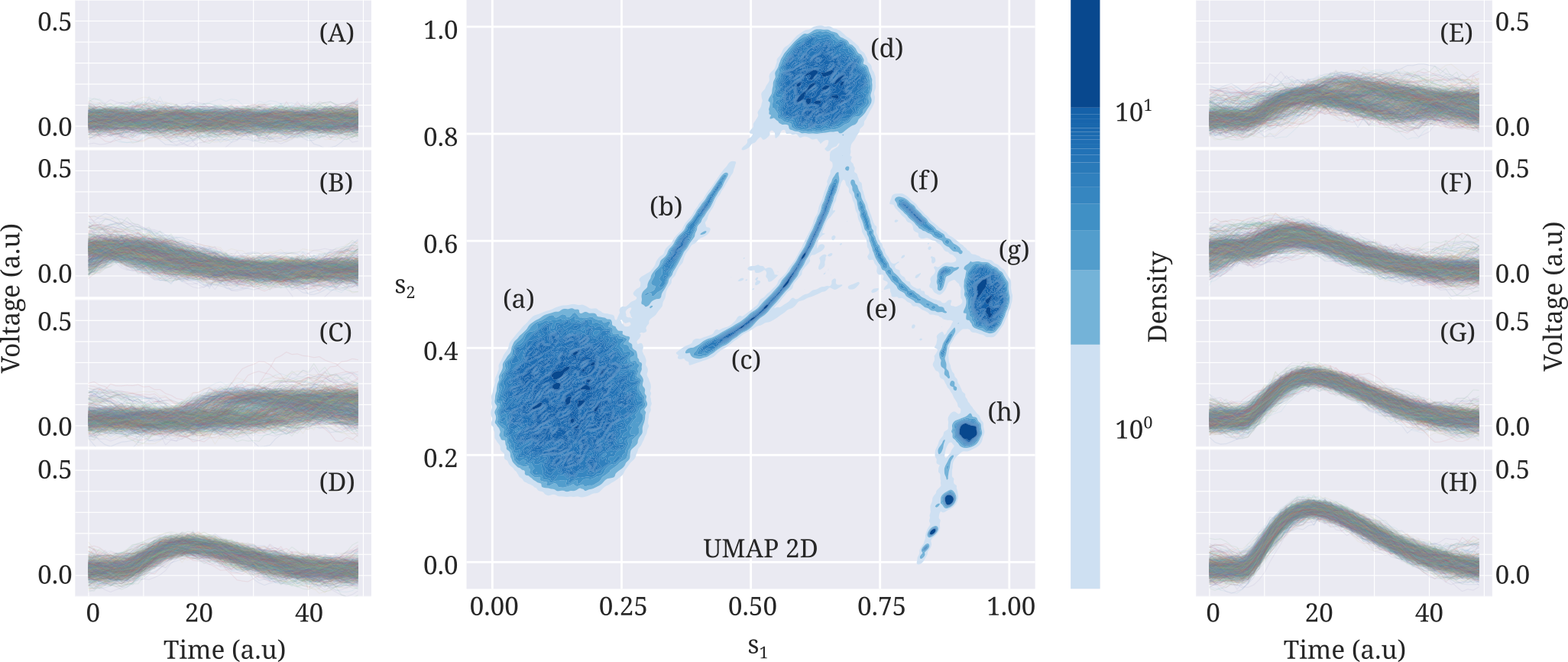}
	\caption{In the centre, we present a low dimensional representation using UMAP of a dataset containing signals from a system with temporally uncorrelated noise. Each cluster in the kernel density estimation is identified using lower case letters, and each graph, identified using the associated upper case letter, represents the signals in each labelled clusters. \textbf{(a)}, \textbf{(d)}, \textbf{(g)}, and \textbf{(h)} give the temporally correlated photon numbers 0 to 3. \textbf{(b)} and \textbf{(c)} are associated with uncorrelated signals, with zero photons correlated before and after the trigger time. \textbf{(e)} and \textbf{(f)} are single photons at the trigger time and uncorrelated signals after and before the trigger. The remaining samples in cluster~\textbf{(i)} are associated with the higher photon numbers and remaining noise photons.}
	\label{fig:noiseUMAP}
\end{figure*}
Using a second dimension, the uncorrelated light becomes distinguishable as shown in Fig.~\ref{fig:noiseUMAP}. In this space, it is not only easier to interpret the proportion of uncorrelated light, but it is also possible to remove these outliers by carefully selecting the latent space regions associated with correlated light.

We also noticed that methods that preserve local structures tend to create clearer clusters for noise structures, facilitating the clustering task. This effect is seen in Fig.~\ref{fig:noiseUMAP} where the uncorrelated noise is found on curve structures and photon numbers in circular shapes. If we look closely at the content of these clusters, we see that it is possible to identify signals of uncorrelated single photons before the trigger time (cluster~\ref{fig:noiseUMAP}b) and after the trigger time (cluster~\ref{fig:noiseUMAP}c). Similarly, we find uncorrelated single photons combined with correlated single photons in clusters~\ref{fig:noiseUMAP}e and \ref{fig:noiseUMAP}f. In clusters~\ref{fig:noiseUMAP}a, \ref{fig:noiseUMAP}d, \ref{fig:noiseUMAP}g, and \ref{fig:noiseUMAP}h we find the standard photon numbers 0 to 3 without uncorrelated light. Similar analysis could be done using more traditional methods like PCA, however the clustering becomes significantly harder. This lack of cluster structure is visually demonstrated in Fig.~\ref{fig:noisePCA} where the uncorrelated light becomes a broadening of the temporally correlated photon numbers.

The remaining samples in cluster~\ref{fig:noiseUMAP}i are associated with the higher photon numbers and remaining noise photons. Since a quasi-thermal light source is used, these signals only represent a small portion of the dataset.

We note that the Gaussian Mixture Model is not as effective in clustering noise features, especially considering a photon-number embedding from UMAP. We found that methods like HDBSCAN, which is a hierarchical density-based clustering technique, are well-suited for UMAP embedding~\cite{Malzer_2020}. This technique has the main advantage of working on clusters that do not follow a Gaussian structure, which is adequate for noise clusters that can have a variety of shapes.

\subsection{Correlated Photons}
To further quantify the impact of noise photons and demonstrate that the clusters identified by UMAP correspond to real physical features, we examine how photon-number assignment influences the reconstructed photon-number distributions of the correlated channels in the Noise dataset. In this system, we expect thermal photon statistics on each individual channel, with second-order coherence approaching $g^{(2)}(0)=2$, along with an added noise contribution from pump leakage. The resulting statistics should therefore lie between Poissonian and thermal behaviour, yielding a $g^{(2)}(0)$ value between 1 and 2, where 1 corresponds to purely Poissonian light and 2 to purely thermal light. As a baseline, we begin by analyzing data obtained when we optimize the pump filtering, i.e. minimize the amount of temporally uncorrelated noise, we obtain values of $g^{(2)}(0)$ of approximately $1.848 \pm 0.008$  for the first channel and $1.879 \pm 0.004$ for the second. The uncertainties reflect the spread in values obtained using different numerical methods. Given the low mean photon number ($\bar{n} \sim 1.01$), the computed values provide a reliable approximation of the true $g^{(2)}(0)$ parameter for a thermal light source, largely independent of the specific numerical method employed. 

Next, we analyze data obtained with increased noise by reducing the pump filtering (which is the data found in the Noise dataset). Using one-dimensional PCA, we obtain $g^{(2)}(0)$ values of 1.514 and 1.592 for the two channels, respectively. A comparable result is observed with the integrated area method, yielding values of 1.431 and 1.459. However, upon using the extra features obtained by UMAP we can reject events associated with noise and thus correct some potential dark counts. As a result, the $g^{(2)}(0)$ values increase to 1.695 and 1.699, bringing them closer to the expected thermal regime. This emphasizes the effectiveness of UMAP-based preprocessing in isolating and mitigating noise. 

In parallel, we estimate the noise reduction factor ($\mathrm{NRF}$) of the clean source, defined as:
\begin{align}
  \mathrm{NRF} = \frac{\mathrm{Var}(n_1 - n_2)}{\langle n_1 + n_2 \rangle},
\end{align}
where $n_1$ and $n_2$ denote the photon numbers in the two channels~\cite{vaidyaBroadbandQuadraturesqueezedVacuum2020d}. For comparison, a Poissonian source yields $\mathrm{NRF} = 1$, whereas twin beams states exhibit $\mathrm{NRF} < 1$, with values increasing in the presence of loss or noise. Across our numerical methods (area, PCA, and UMAP), we compute a baseline value of $\mathrm{NRF} = 0.772 \pm 0.005$.

For the $\mathrm{NRF}$, we compute values of 0.826 (area), 0.810 (PCA), and 0.811 (UMAP), showing only a modest increase from the baseline. Notably, the $\mathrm{NRF}$ shows limited sensitivity to the choice of numerical method. We attribute this to the low system efficiency, characterized by a transmission coefficient of approximately $\eta \approx 22\%$. Since the $\mathrm{NRF}$ scales as $1 - \eta$, it approaches 1 in low-efficiency regimes, where it becomes less sensitive to noise. This contrasts with $g^{(2)}(0)$, which remains loss-insensitive. A simulation illustrating this effect is available in a public repository~\cite{repo}.

\begin{figure}[htbp]
  \centering
  \includegraphics[width=.95\linewidth]{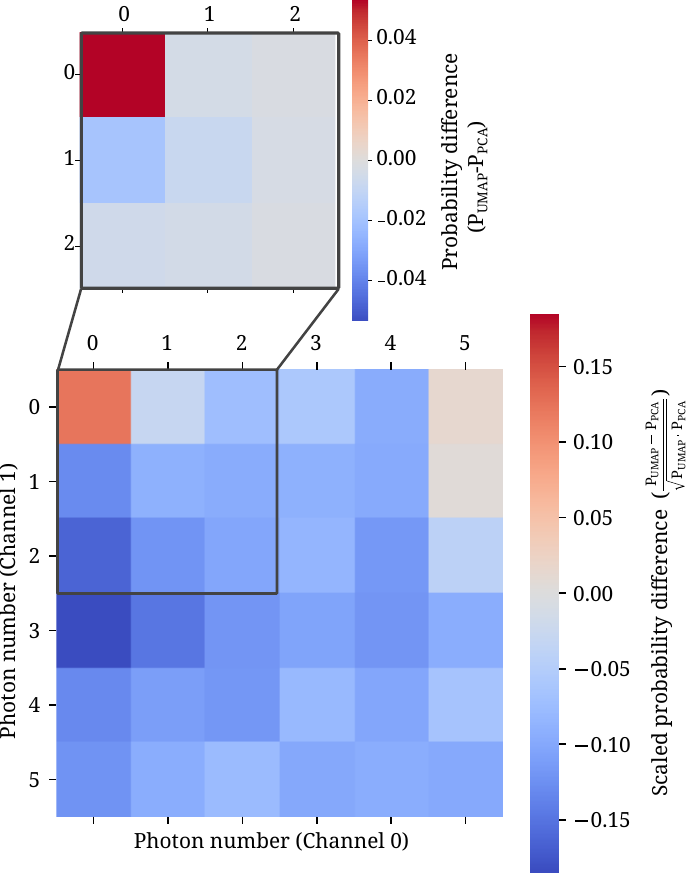}
  \caption{Scaled difference between the joint photon-number distributions computed using UMAP (2D) and PCA (1D) on the Noise dataset (light source described in Sec.~\ref{sec:dataset}). For reference, the unscaled joint distribution is shown for the first few photon-number states.}
  \label{fig:jointDistSub}
\end{figure}

To further illustrate the effect of photon-number misclassification, Fig.~\ref{fig:jointDistSub} presents the scaled difference between the joint photon-number distributions derived from PCA and UMAP. The most prominent deviation stems from PCA misclassifying 0-photon events approximately 4\% of the time due to noise influence, reinforcing the advantage of UMAP for denoising and accurate state reconstruction. This effect is also reflected in the predicted mean photon number across the different techniques, as temporally uncorrelated noise primarily decreases the number of predicted 0-photon events. We observe a decreasing trend in the estimated mean photon number that correlates with each method's ability to correctly identify and reject noise. The area method yields a mean photon number of 0.697, as it effectively integrates the total signal over the full time interval. PCA, which partially filters traces based on temporal features, reduces the mean to 0.637. UMAP, with its stronger noise discrimination capabilities, further decreases the estimated mean to 0.566. 

\subsection{Impact of Gaussian Mixture Model}
\begin{figure*}[t]
 	\centering
 	\begin{subfigure}[b]{0.45\textwidth}
     	\centering
     	\includegraphics[width=\textwidth]{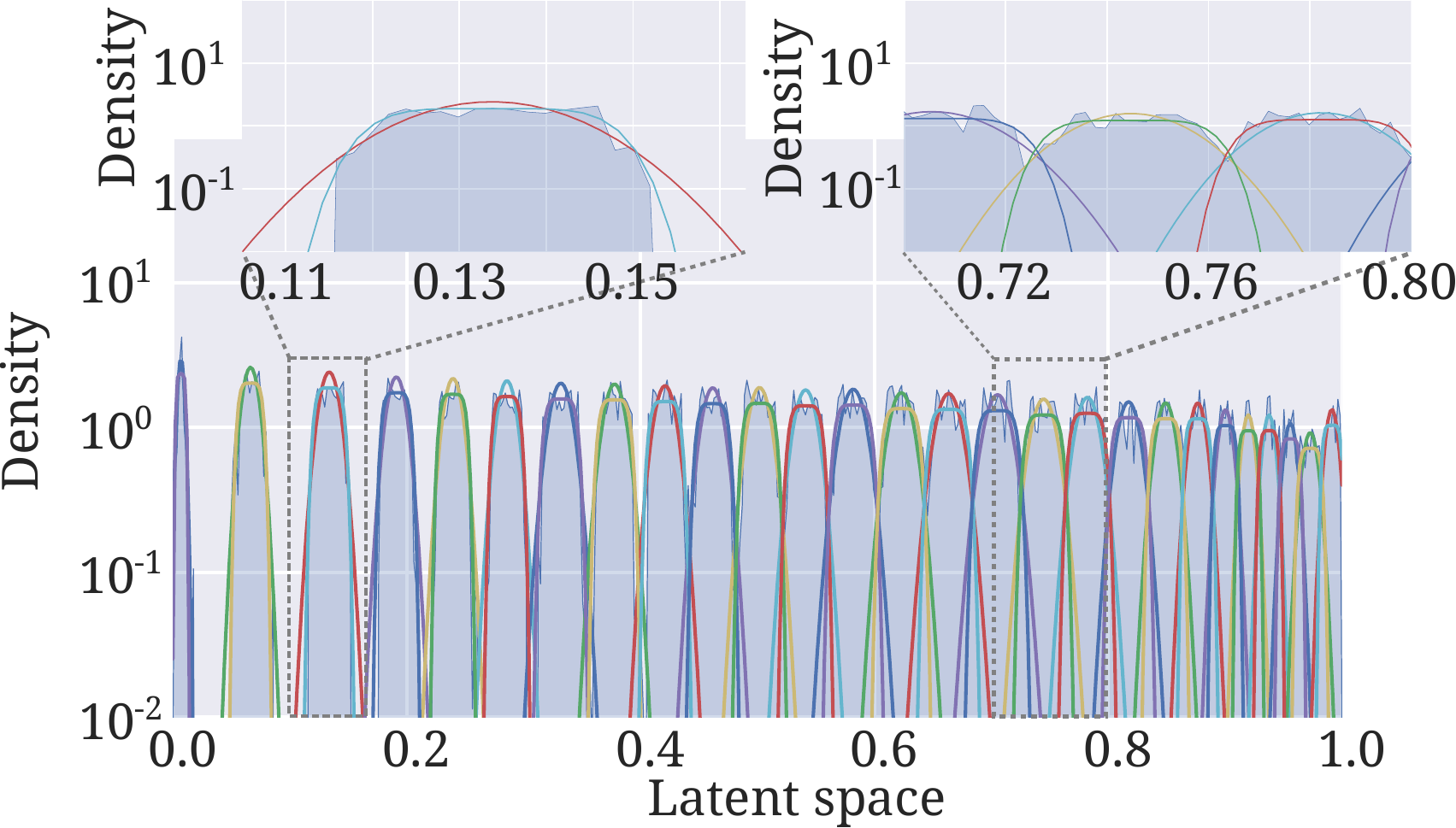}
     	\caption{Gaussian and generalized Gaussian fit over the kernel density estimation of the t-SNE 1D embedding, where the broader distributions are the standard Gaussians.}
     	\label{fig:GGF}
 	\end{subfigure}
 	\hfill
 	\begin{subfigure}[b]{0.45\textwidth}
     	\centering
     	\includegraphics[width=\textwidth]{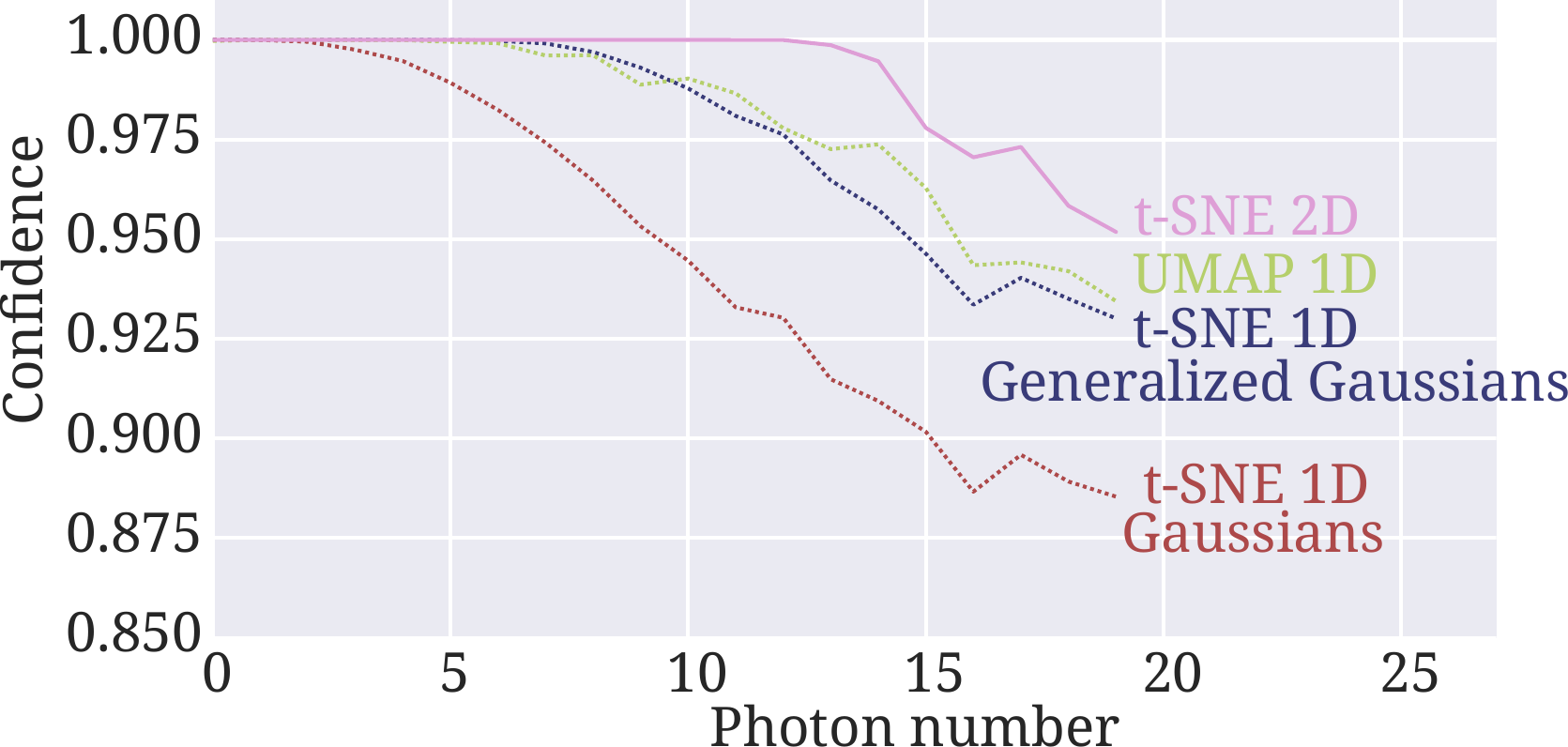}
     	\caption{Confidence associated with the one-dimensional embedding from t-SNE using standard and generalized Gaussian functions to describe the clusters.}
     	\label{fig:Confidence_GGF}
 	\end{subfigure}
	\caption{Impact of using a generalized Gaussian function to estimate the clusters generated by t-SNE.}
	\label{fig:three graphs}
\end{figure*}

One-dimensional results for t-SNE offer clusters that follow top-hat-like distributions, cf. Fig.~\ref{fig:lowDim}. This feature decreases the confidence results, but not the actual potential clustering over this embedding. For a more accurate representation of t-SNE clusters, we use a generalized Gaussian distribution to represent the probability density of each cluster defined as
\begin{align}
	p(s|n)=\frac{\beta}{2\zeta_n\Gamma(1/\beta)}\exp\left[-\left| \frac{s-\mu_n}{\zeta_n}\right|^{\beta}\right],
\end{align}
with
\begin{align}
	\zeta_n^2 = \frac{\sigma_n^2 \Gamma(1/\beta)}{\Gamma(3/\beta)}.
\end{align}
In these equations, $\mu_n$, $\sigma_n^2$, and $\Gamma$ are respectively the mean and variance of a given photon-number cluster and the Gamma function. In Fig.~\ref{fig:GGF} we present a qualitative representation of the fit quality of t-SNE embedding using the standard and generalized Gaussian functions.
In Fig.~\ref{fig:GGF} we see that the generalized Gaussian distribution is a better estimation of the density inside the latent space. The small tail reduces greatly the overlap of probability density functions, which increases the computed confidence.
The new values of confidence are plotted in Fig.~\ref{fig:Confidence_GGF} where we observe a significant increase in the confidence, reaching values similar to one-dimensional UMAP.

\subsection{Potential Implementations}
Based on the benchmarks, the dimensionality reduction techniques that focus on local structure preservation offer the best low-dimensional representation of the transition-edge sensor signals.

These methods provide high cluster separation and follow the expected distribution to a degree unmatched by other techniques. For these reasons, t-SNE and UMAP are effective methods for applications that do not involve frequently adding new samples to their dataset and require high accuracy. The existence of open-source platforms like UMAP-learn~\cite{umap_nodate} and Scikit-Learn~\cite{scikit-learn_nodate} that offer complete and optimized implementations of these methods facilitates its usability. The number of user parameters necessary to use these methods is also very small, which makes them ideal for experts and non-experts. We note that the operation complexity scaling of UMAP is much more advantageous when compared to t-SNE, it is therefore more efficient to use UMAP since they both have similar performances.

Considering the previous performance results, neural networks  (5 linear layers of 300 neurons) can offer a trustworthy and interpretable low-dimensional representation of the TES traces. The condition necessary for this method to be accurate is to provide a balanced dataset containing the range of photon numbers we want to detect. It is essential to understand that the network cannot predict photon numbers outside the trained range, since it never learned an embedding for these signals. The training data restricts the learned transformation. Moreover, our results suggest that a small neural network implemented in a Field Programmable Gate Array (FPGA)~\cite{lingala_fpga_2022} could replace currently used methods like trace area and PCA~\cite{moraisPreciselyDeterminingPhotonnumber2022a} to process the TES traces directly. With this type of hardware we believe real-time processing can be achieved considering TESs have a dead time of a few microseconds and knowing our CPU implementation can process a TES signal of 200 points in $4.9\mu s$. This value is obtained using a laptop with a clock speed of 3.2 GHz, 8 cores and 16 threads.

We emphasized using a close-to-uniform distribution to train the network, since it becomes equally optimized for every class (photon number). Following the example used to benchmark the different methods, the use of a coherent source with a tuneable mean photon number is more than sufficient to create a balanced dataset. It is therefore possible to create suitable conditions only using a laser and tuneable attenuation. We could also imagine using a high mean photon number thermal source, depending on the available equipment.

\subsection{Future Work}
A one-dimensional embedding is optimal for experimental systems where the number of possible outliers is limited since the clustering task becomes simplified. To improve on this work, we hypothesize that there is a solution in one dimension that can reach the confidence values of two-dimensional UMAP and t-SNE. To address this problem, we could enhance our understanding by examining the relationship between the dimensionality reduction process and clustering. Additionally, it may be possible to strengthen the representation of photon numbers while minimizing the space allocated to noise features.

While testing the different methods, the clustering step (Gaussian Mixture Model) was particularly sensitive to the initialization process. Often some manual adjustments had to be done to guarantee the quality of the results. To further improve the quality and robustness of photon-number classification, future work could explore clustering techniques that may be better aligned with the novel methods introduced in this study. This way it could be possible to completely automate the photon-number classification process even for low visibility clusters.

Coming back to the use of methods that preserve local structures, we believe that using methods like UMAP can enable the use of TESs on temporally uncorrelated light, making it a useful tool to remove noise in a variety of cases. Also, this feature can be exploited to characterize photon statistics of continuous-wave sources where no time trigger can be used. Existing work on the topic~\cite{lee_multi-pulse_2018} uses a different approach to this problem, making it difficult to compare. However, the methods we describe make this task simple to implement for a wide variety of cases, since it is invariant to the combinations of photon events inside a single signal. In other words, traces associated with exotic scenarios, for example a single photon trace slightly overlapped by a two-photon trace, should have its position inside the latent space making it distinguishable. This task is also well suited for neural networks since they can be designed to be shift-invariant, meaning that similar structures, independent of their position, could be clustered.  

Accessing the ground truth for photon-number classification remains a significant challenge, and further validation is needed to robustly assess the performance of the proposed methods. To this end, we leverage the joint probability distribution of twin beams to evaluate the impact of dimensionality reduction techniques (see Sec.~\ref{sec:outlier}). This distribution serves as an intuitive diagnostic tool, analogous to a confusion matrix commonly used in classification tasks. In our current implementation, generating twin beams with high mean photon numbers and strongly correlated channels is not feasible. Nevertheless, the distribution proves useful for assessing the filtering capabilities of the methods under study. A future experiment with higher photon numbers, lower loss, and reduced noise would offer deeper insight into the potential of this numerical processing. Given these considerations, we anticipate broadening effects introduced by the numerical techniques; thus, the width of the joint probability distribution becomes a practical experimental observable to quantify their performance.

\section{Conclusion}\label{sec:conclusion}
Nonlinear methods like t-SNE and UMAP that aim to preserve local data structures offer better resolution over photon numbers in the case of transition-edge sensor signals compared to previously used techniques like signal area and PCA. These methods can be used directly to replace currently used methods, with the caveat that they cannot predict new samples without computing the entire dataset. However, with a large dataset ($u=550\,000$ samples), we demonstrate the potential of neural networks that recreate the embedding of t-SNE and UMAP. These models remain simple and could be further explored, offering a promising direction for future research. Enhancing the generalization capabilities of these models could enable their application in real-time photon-number resolution.

This enhanced photon-number resolution has direct implications for the preparation of non-classical states of light, such as cat states, GKP states, and magic states~\cite{walschaersNonGaussianQuantumStates2021}, which require precise heralding and photon-number discrimination. In particular, higher-resolution detection allows for more accurate post-selection, improved state fidelity, and greater control over resource-state generation, which are all critical for fault-tolerant quantum computation and quantum error correction.

Beyond TES devices, the techniques explored in this work hold promise for enhancing the performance of other single-photon detectors, such as SNSPDs. For instance, principal component analysis (PCA) has shown potential in processing SNSPD signals~\cite{schapeler_how_2023,divochiy_superconducting_2008,sauerResolvingPhotonNumbers2023c}, highlighting the versatility of these approaches across photon-detection technologies.

During the publication process, we became aware of an independently conducted work that similarly explores the use of advanced signal classification techniques to improve TES detection performance~\cite{liBoostingPhotonnumberresolvedDetection2025}. We acknowledge this complementary work, which highlights the growing interest in applying machine learning to enhance the speed and efficiency of photon-number resolving detectors.

All the numerical methods and data discussed in this document are available in Ref.~\cite{repo,Dalbec-Constant2024-ap,Gerrits2024-ap}.

\section*{Acknowledgements}
N.D.-C. and N.Q. acknowledge support from the Ministère de l'Économie et de l'Innovation du Québec, the Natural Sciences and Engineering Research Council Canada, Photonique Quantique Québec, and thank S. Montes-Valencia, J. Mart\'inez-Cifuentes and A. Boon for valuable discussions. We also thank Z. Levine and S. Glancy for their careful feedback on our manuscript.

\bibliography{References}
\end{document}